\documentclass[aps,prd,longbibliography,nofootinbib,amsthm,amsmath,amssymb,amsfonts,notitlepage]{revtex4-1}
\usepackage[utf8]{inputenc}
\usepackage[T1]{fontenc}
\usepackage[english]{babel}
\usepackage{graphicx}
\usepackage{float}
\usepackage{cleveref}
\usepackage{xcolor}
\usepackage{color}
\usepackage{tikz}
\usepackage{multirow}
\usepackage{appendix}
\usepackage{slashed}
\usetikzlibrary{arrows}
\usepackage{braket}
\usepackage[section]{placeins}

\date{\today}

\begin{document}
	\title{ Momentum space oscillation properties of vortex states collision}
	\author{Pengcheng Zhao}
	\email{zhaopch5@mail2.sysu.edu.cn}
	
	\affiliation{School of Physics and Astronomy, Sun Yat-sen University, 519082 Zhuhai, China}

	\begin{abstract}
		A qualitative calculation and discussion of two vortex states collisions are given in the scalar $\phi ^4$ model.
		Three kinds of vortex states --- Bessel, general monochromatic, and Laguerre-Gaussian vortex states --- are considered.
		It is found that the total final momentum distribution in collision  of physical vortex states displays general topological structures, which depend on the initial vortex states' topological charges, which are proportional to the orbital angular momenta.
		This peculiar matching provides a novel observable, the topological number of momentum distribution, and it may represent a new fascinating research direction in particle physics.
		We also find that the situation when the angular momenta of the two colliding Laguerre-Gaussian states combine to zero can be recognized by the total final momentum distribution close to the collision axis.
		Both features can be used to measure the orbital angular momentum of vortex states.
		
	\end{abstract}
	\maketitle
	\section{Introduction}
	In vast majority of cases, it is enough to calculate particle interactions using the plane wave approximation.
	Nevertheless, many studies have shown that non-plane wave nature of particles plays an important role in physical processes.
	For example, the so-called MD effect, which reveals the importance of the impact parameter, can only be explained when describing particles as wave packets rather than plane waves \cite{Large impact parameter cut-off in the process e+e- to e+e-gamma, Deviation from standard QED at large distances: Influence of transverse dimensions of colliding
		beams on bremsstrahlung}.
	The so-called ``beam size effect'' reveals that size of particles cannot be neglected when it is comparable to or smaller than the particle Compton wavelength \cite{Physical mechanism of the linear beam size effect at colliders, Processes in the T channel singularity in the physical region: finite beam sizes make cross sections finite, Scattering of wave packets on atoms in the Born approximation, Loss of wave-packet coherence in ion-atom collisions}.
	Also, vortex particle collisions exhibits peculiar features which cannot be found in the plane wave approximation \cite{Theory and applications of free-electron vortex states, Promises and challenges of high-energy vortex states collisions}.
	In this paper, we will demonstrate yet another special feature arising in two vortex states collisions which has no counterpart for plane wave scattering.
	
	Vortex states have quantized intrinsic angular momentum which is different from spin.
	They are characterized by a spiral phase front which is described by the extra phase factor $e^{i \ell \varphi}$, 
	where $\ell$ is called the topological charge and is proportional to the orbital angular momentum.
	Several kinds of particles prepared in such a state have been experimentally produced: photons \cite{Helical-wavefront laser beams produced with a spiral phaseplate}, electrons \cite{Generation of electron beams carrying orbital angular momentum}, neutrons \cite{Controlling neutron orbital angular momentum}, and small atoms \cite{Vortex beams of atoms and molecules}.
	They have found applications in fields such as manipulation of atoms \cite{Mechanical equivalence of spin and orbital angular momentum of light: an optical spanner, Dynamic holographic optical tweezers, Microoptomechanical pumps assembled and driven by holographic optical vortex arrays}, quantum communication \cite{Quantum Correlations in Optical Angle-Orbital Angular Momentum Variables, Terabit free-space data transmission employing orbital angular momentum multiplexing, Chiral specific electron vortex beam spectroscopy, Probing the electromagnetic response of dielectric antennas by vortex electron beams}, microscopy  \cite{Local orbital angular momentum revealed by spiral phase plate imaging in transmission electron microscopy} and so on.
	Along with generation of vortex particles, the study of particle interactions involving vortex states attracts a lot of attention \cite{Promises and challenges of high-energy vortex states collisions}.
	Many processes involving vortex particles have been studied, for example, vortex particle propagation in the electromagnetic fields \cite{Production of twisted particles in magnetic fields, Evolution of an accelerated charged vortex particle in an inhomogeneous magnetic lens, Transmission of vortex electrons through a solenoid}, vortex particle scattering by a target \cite{Study of highly relativistic vortex electron beams by atomic scattering, Scattering of twisted relativistic electrons by atoms, Observability of the superkick effect within a quantum-field-theoretical approach}, decay of the vortex muons \cite{Decay of the vortex muon, Zn symmetry in the vortex muon decay}, photo-induced processes  initiated by vortex photons \cite{Recoil Momentum Effects in Quantum Processes Induced by Twisted Photons, Delta baryon photoproduction with twisted photons, Threshold effects in high-energy vortex state collisions} and so on .
	There are also several studies focusing on collision of two vortex particles \cite{Colliding particles carrying non-zero orbital angular momentum, Measuring the phase of the scattering amplitude with vortex beams, Elastic scattering of vortex electrons provides direct access to the Coulomb phase}.
	A key characteristic of two vortex particle collisions is a peculiar oscillatory dependence of the differential cross section on the transverse momentum.
	This sort of oscillation depends on the topological charges of the initial vortex states.
	However so far this dependence has only been discussed with Bessel vortex states, which are not physical because they cannot be normalized in the transverse space.
	A physical vortex state can have longitudinal momentum uncertainty and an energy spread.
	Thus, the physical vortex states collision may result in special longitudinal momentum or energy distributions.
	The dependence of these distributions on the topological charge is not immediately clear.
	
	In this paper, we clarify this issue. To this end, we study collisions of two vortex particles
	in three different cases, when they are prepared in Bessel vortex state, in a general monochromatic vortex state, and in the Laguerre-Gaussian vortex state.
	Transverse, longitudinal and energy distributions are calculated. 
	We compare the three cases and find general topological features which depend on the topological charges of the two initial vortex states.
	To keep the discussion clear, we consider collisions in the scalar $\phi ^4$ theory.
	
	The paper is organized as follows.
	In section \ref{section two}, we present calculations of the transition amplitude ${\cal S}$ for the three different cases.
	In section \ref{section three}, we discuss and compare the resulting total final momentum distributions.
	In section \ref{section four}, we draw conclusions.
	In this paper, all the three-dimentional vectors are expressed using bold letters.
	
	\section{collision of two vortex states}\label{section two}
	In this study we only consider central collisions of vortex states that carry quantized orbital angular momenta along their propagation direction, which we label as $z$ axis.
	This means that the vortex states are eigenstates of the $z$-component of the angular momentum operator.
	Central collisions of these eigenstates must produce the total final momentum distribution which is rotational symmetric around the axis $z$, if we measure the final states in 3-momentum basis, i.e. plane wave basis.
	In this kind of distribution, we have the freedom to choose the transverse axes; we select axis $x$ along the transverse part of total final momentum.
	\subsection{Bessel vortex collision}
	\begin{figure}[!h]
		\centering
		\includegraphics[width=0.7\textwidth]{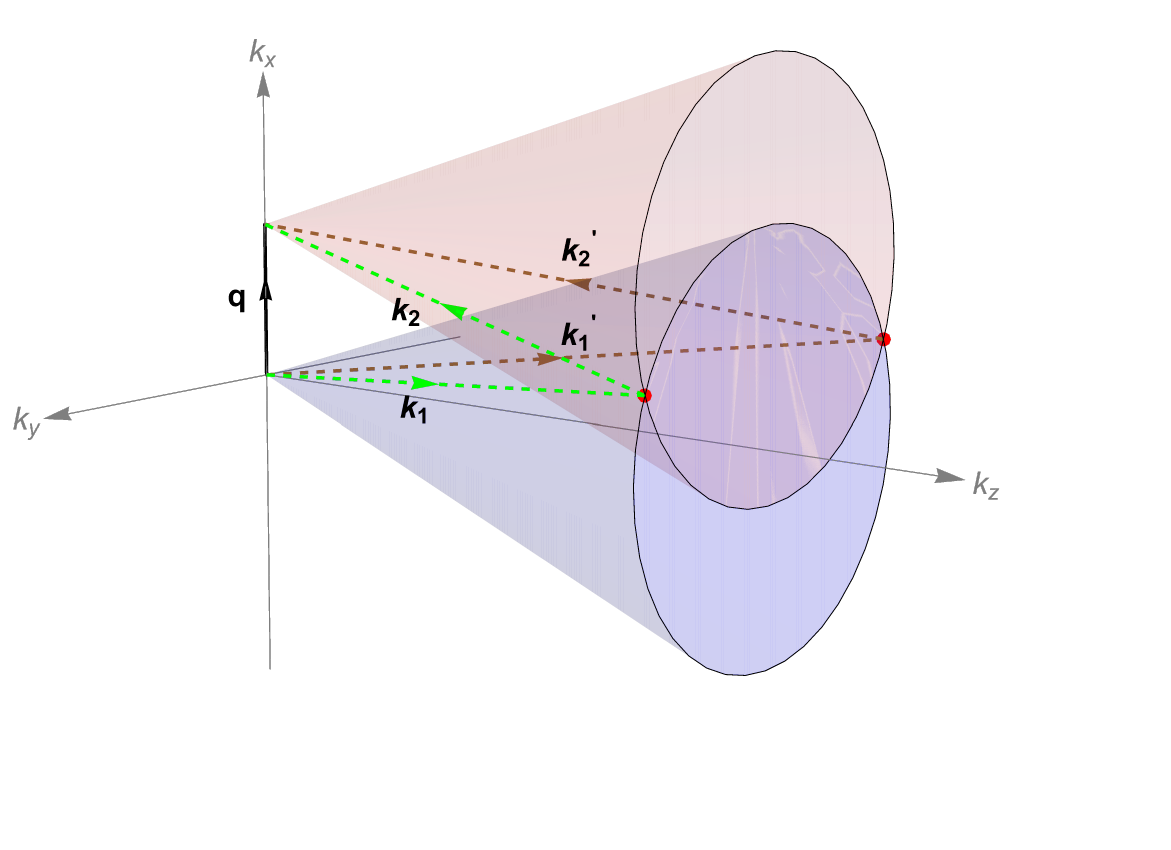}
		\caption{Interference scheme for Bessel vortex states collision. The bases of the two momentum cones intersect at two points which give the same total final momentum $\bold q$. The two interfering pairs of plane wave components are shown 
			as  $({\bold k_1}, {\bold k_2})$ and $({\bold k_1'}, {\bold k_2'})$.}\label{two points}
	\end{figure}
	
	A Bessel vortex state is characterized by three quantum numbers: the energy $E$, the longitudinal momentum $p_z$, and topological charge $\ell$, which can only be integer.
	The Bessel state cannot be normalized in the transverse plane.
	The momentum space distribution of this state is just a circle orthogonal to the axis $z$.
	Momenta of all the plane wave components inside the Bessel state form a cone whose base is just this circle.
	The state propagating along $z$ direction reads
	\begin{eqnarray}
		\ket {\psi^B_{E,p_z,\ell}(\bold x,\,t)}&=&\int \frac{d^3\bold k}{(2\pi )^3}a^B_{p_z \ell}(\bold k)\,{\rm{exp}}(i \bold k \bold x-i Et),\nonumber \\
		a^B_{p_z \ell}(\bold k)&=&N^B\delta (k_{\perp}-\kappa)\delta (k_z-p_z)\,\rm{exp}(i \ell \varphi _k),
	\end{eqnarray}
	where $E=\sqrt{\bold k^2+m^2}$ is the energy of the plane wave with momentum $\bold k$ and mass $m$, $\kappa=\sqrt {\bold k^2-p_z ^2}$ is   modulus of the transverse momentum, and $\varphi_k$ is the azimuthal angle of this plane wave. $N^B$ is the normalization coefficient, but we will not need its explicit expression.
	
	Being different from the plane wave particle interactions, collisions of two Bessel vortex states show some new features that can be observed, including non-constant total transverse momentum and the oscillatory transverse momentum distribution \cite{Elastic scattering of vortex electrons provides direct
		access to the Coulomb phase, Promises and challenges of high-energy vortex states collisions}.
	The new distributions represent new physical observables which are absent in traditional collision kinematics.
	The momnetum space oscillations appear because of interference between different plane wave components present in the initial vortex states.
	For non-plane wave state collisions, ``interference'' means that we can find multiple initial plane wave combinations which lead to the same final state.
	For Bessel vortex state collisions, this interference involves only two isolated points in momentum space, as shown in Fig. \ref{two points}.
	In this figure, $\bold q$ is total final momentum, $\bold k_1$ and $\bold k_2$ label the momenta of plane wave components of the two initial vortex states. The upper black circle is momentum distribution of $(\bold q-\bold k_2)$ and the lower black circle is the momentum distribution of $\bold k_1$.
	The two circles intersect at two red points, which satisfy conservation law: $\bold q=\bold k_1+\bold k_2$.
	
	The S-matrix element for a process 
	``Two Bessel vortex states to two plane wave states'' was studied in reference \cite{Colliding particles carrying non-zero orbital angular momentum} and the result is
	\begin{eqnarray} \label{Bessel collision}
		{\cal S}(B_1+B_2\rightarrow P_3+P_4)&\propto &\int d^3\bold k_1 d^3 \bold k_2 a^B_{p _{1z}\ell _1}(\bold k_1) a^B_{p_{2z}\ell _2}(\bold k_2){\cal M}\,\delta ^3(\bold k_1+\bold k_2-\bold q)\delta(E_1+E_2-E_q)\nonumber\\
		&\propto &\frac{({\cal M}^++{\cal M}^-)}{\Delta}\,\delta (k_{1z}+k_{2z}-q_z)\delta(E_1+E_2-E_q)\nonumber\\
		&\propto&\frac{\cos (\ell _1 \varphi _1+\ell _2 \varphi _2)}{\Delta}\,\delta (k_{1z}+k_{2z}-q_z)\delta(E_1+E_2-E_q),
	\end{eqnarray}
	where subscript $1,\,2$ represent the two initial states respectively, $E_q$ is the total energy of final state, and the amplitudes ${\cal M}^+$ and ${\cal M}^-$ correspond to the two interfering paths shown in Fig. \ref{two points}, 
	and
	\[\Delta=\sqrt{(\kappa_1+\kappa_2+q_{\perp})(\kappa_1-\kappa_2+q_{\perp})(-\kappa_1+\kappa_2+q_{\perp})(\kappa_1+\kappa_2-q_{\perp})}\]  is the area of the triangle, lying in the transverse plane, formed by $(\kappa_1,\,\kappa_2,\,q_{\perp})$.
	Here we have chosen the azimuthal angle of $\bold q$ to be $\varphi_q=0$, which means that $\bold q$ lies in the $xOz$ plane. 
	We also define the topological charges according to one same $z$ axis, even though the two particles are counterpropagating.
	Noting that it is a point-like interaction for a $2\rightarrow 2$ process in scalar $\phi ^4$ theory if we only consider tree level diagrams.
	Hence, calculation from second line of Eq. (\ref{Bessel collision}) to third line is correct.
	In the third line of Eq. (\ref{Bessel collision}), $(\varphi _i,\,i=1,2)$ is dependent on $q_{\perp}$:
	\begin{eqnarray}
		\varphi_1&=&\arccos ({\frac{\kappa_1^2+q_{\perp}^2-\kappa_2^2}{2q_{\perp}\kappa_1}}),\nonumber\\
		\varphi_2&=&-\arccos ({\frac{\kappa_2^2+q_{\perp}^2-\kappa_1^2}{2q_{\perp}\kappa_2}}).\nonumber
	\end{eqnarray}
	We can also choose $\varphi_1$ to be negative and $\varphi_2$ to be positive.
	The two cases will give same result.
	
	From Eq. (\ref{Bessel collision}) we see that collision of two Bessel vortex states can only generate final states that have the fixed total energy $E_1 + E_1=E_q$ and the fixed longitudinal momentum $k_{1z}+k_{2z}=q_z$. However the total final transverse momentum $q_{\perp}$ is not fixed, but is distributed within the region
	\begin{equation}
		|\kappa_1-\kappa_2|<q_{\perp}<|\kappa_1+\kappa_2|\,,\nonumber
	\end{equation}
	and the scattering matrix amplitude exhibits an oscillatory behavior in this range.
	
	\subsection{General monochromatic vortex collision}
	\begin{figure}[!h]
		\centering
		\includegraphics[width=0.8\textwidth]{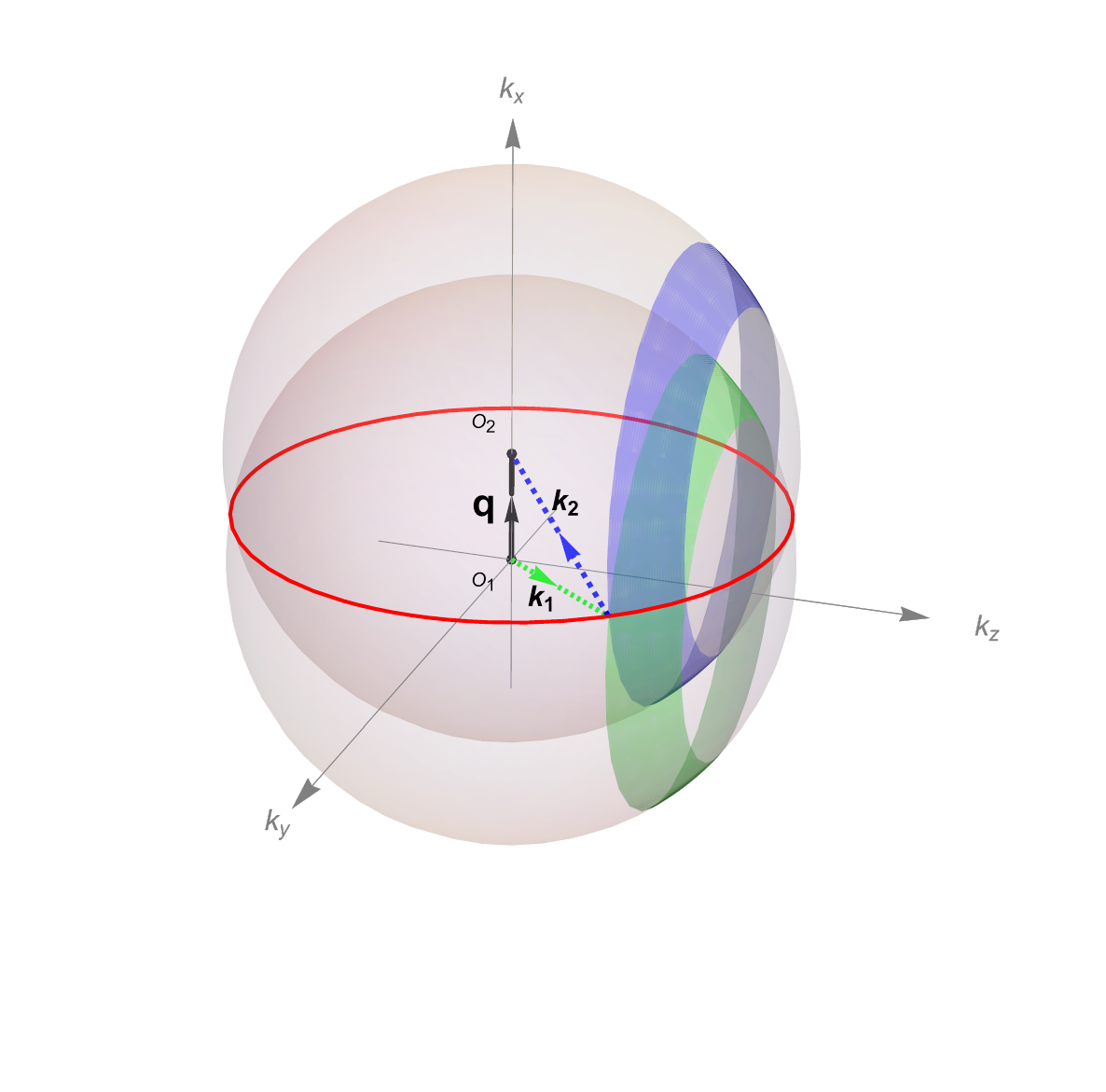}
		\caption{Interference scheme for the general monochromatic vortex states collision. Two momentum spheres intersect at points which result in the same total final momentum $\bold q$. 
			They form the red circle.
			The two girdles in the two spheres show main contribution of the two initial vortex states.  Intersection of the red circle with the two girdles displays main contribution points to the fixed final state.  Here we only show momenta of one contribution point labeling as $(\bold k_1,\,\bold k_2)$ among all the contributions.  }\label{a circle}
	\end{figure}
	
	A general monochromatic vortex state can be written as a superposition of many Bessel vortex states with the same energy.
	It has two quantum numbers: the energy $E$ and the topological charge $\ell$.
	It can be normalized in the transverse plane.
	In momentum space, all the plane wave components have momenta lying on a sphere of radius $k \equiv |{\bold k}|$ 
	and are accompanied by different weighting factors.
	The state propagating along the $z$ direction reads
	\begin{eqnarray}
		\ket {\psi^M_{E,\ell}(\bold x,\,t)}&=&\int \frac{d^3\bold k}{(2\pi )^3}a^M_{\ell}(\bold k)\,{\rm{exp}}(i \bold k \bold x-i Et),\nonumber \\
		a^M_{\ell}(\bold k)&=&N^M f(\theta)\delta (\sqrt {\bold k^2+m^2}-E)\,\rm{exp}(i \ell \varphi _k),
	\end{eqnarray}
	where $\theta$ is the polar angle of the momentum $\bold k$, $N^M$ is the normalization coefficient of the state.
	Note that $f(\theta = 0) = f(\theta = \pi) = 0$ to avoid phase singularity.
	In realistic situations, $f$ should be concentrated in a narrow region of $\theta$.
	For the numerical calculations, we use 
	\begin{eqnarray}\label{weight fucntion}
		f(\theta)=\sin \theta\,{\rm {exp}}\left[-\frac{(\theta-\theta_0)^2}{\sigma _0^2}\right],
	\end{eqnarray}
	where $\theta_0$ and $\sigma_0$ are small constants.
	
	For a fixed total final 4-momentum, there are many plane wave combinations that satisfy 
	the energy-momentum conservation law and lead to the same final state.
	A schematic analysis of the contributions in momentum space is shown in Fig. \ref{a circle}.
	The red circle, which is the intersection of the two spheres, is perpendicular to the total final momentum $\bold q$ 
	and displays all points which can lead to the same final state.
	One sphere shows $\bold k_1$ distribution and another displays $(\bold q-\bold k_2)$ distribution, where $(\bold k_1,\,\bold k_2)$ represent momentum components in the two initial vortex states.
	Only intersection points are selected by the conservation law.
	Since momenta of the initial vortex states are concentrated within certain bands as shown in Eq. (\ref{weight fucntion}), the main contribution to the production amplitude comes from the intersection of the red circle with two colored bands.
	
	The ${\cal S}$-matrix element for the process ``Two monochromatic vortex states to two plane wave states'' is
	\begin{eqnarray} \label{Monochromatic collision}
		{\cal S}(M_1+M_2\rightarrow P_3+P_4)&\propto &\int d^3\bold k_1 d^3 \bold k_2 a^M_{\ell _1}(\bold k_1) a^M_{\ell _2}(\bold k_2){\cal M}\,\delta ^{(3)}(\bold k_1+\bold k_2-\bold q)\delta(E_1+E_2-E_q)\nonumber\\
		&\propto &\int _0^{\pi} d\hat \varphi_1\frac{k_1k_2}{q}f_1(\theta_1)f_2(\theta_2)  \cos(\ell_1 \varphi_1+\ell_2 \varphi_2){\cal M}\,\delta(E_1+E_2-E_q),
	\end{eqnarray}
	where $k_1$, $k_2$ and $q$ are modulus of momentum $\bold k_1$, $\bold k_2$ and $\bold q$.
	$\theta$ is polar angle and $\varphi $ is azimuthal angle in momentum space.
	Note that we use two coordinate systems in Eq.~(\ref{Monochromatic collision}).
	One of them defines $z$ axis along propagation direction.
	Another one defines $z$ axis along total momentum $\bold q$ and angles in this system are labeled with hats.
	In the second line of Eq. (\ref{Monochromatic collision}), $(\varphi _i,\theta_i,\,i=1,2)$ depend on $(q,\theta _q,\hat \varphi_i)$:
	\begin{eqnarray}
		\varphi_i&=&\arccos ({\frac{\cos \hat\theta _i\sin \theta _q+\sin \hat\theta_i\cos \theta_q \cos \hat\varphi _i}{\sqrt{1-(\cos \hat\theta _i \cos \theta _q-\sin \hat\theta _i\sin \theta _q\cos \hat\varphi _i)^2}}}),\nonumber\\
		\theta_i&=&\arccos (\cos \hat\theta_i\cos \theta_q-\sin \hat\theta_i\sin \theta_q\cos \hat \varphi_i)\nonumber
	\end{eqnarray}
	where $\theta_q$ is angle between vector $\bold q$ and $z$ axis and
	\begin{eqnarray}
		\hat \varphi_2&=&-\hat \varphi_1,\nonumber\\
		\hat \theta_1&=&\arccos (\frac{k_1^2+q^2-k_2^2}{2qk_1}),\nonumber\\
		\hat \theta_2&=&\arccos(\frac{k_2^2+q^2-k_1^2}{2qk_2}).\nonumber
	\end{eqnarray}
	
	Collisions of two monochromatic vortex states can only generate final states with a fixed energy;
	this is also shown by the energy delta function in Eq.~(\ref{Monochromatic collision}).
	However, the total transverse and longitudinal momentum distributions span non-zero ranges,
	and within these regions, they exhibit oscillations.

	\subsection{Laguerre-Gaussian vortex collision}
	
	\begin{figure}[!h]
		\centering
		\includegraphics[width=0.8\textwidth]{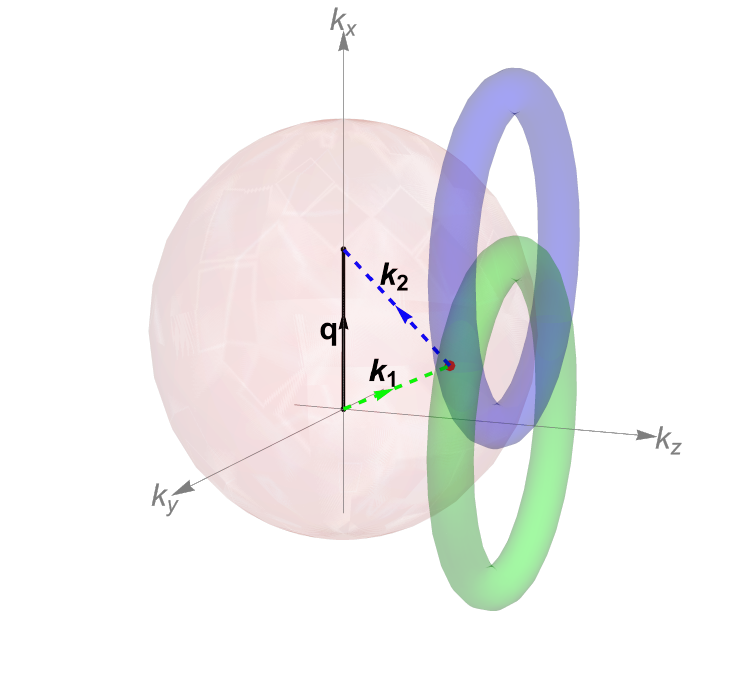}
		\caption{Interference scheme for Laguerre-Gaussian vortex states collision. Surface of an ellipsoid gives all the points who give same total momentum and energy. The two donuts show main contribution of the initial Laguerre-Gaussian states.Intersection of the ellipsoid surface with two donuts gives main contribution to final state. Here we ignore mass of the particles ($m=0$).}\label{an ellipsoid}
	\end{figure}
	
	The Laguerre-Gaussian vortex state can also be written as a superposition of many Bessel vortex states.
	It is not a monochromatic state.
	It can be well normalized on the transverse plane.
	It has two quantum numbers which are the topological charge $\ell$ and principal quantum number $n$, both of which must be integer.
	In momentum space, all the momentum components have contribution to it.
	The state propagating along $z$ direction reads
	\begin{eqnarray}
		\ket {\psi^L_{n,\ell}(\bold x,\,t)}&=&\int \frac{d^3\bold k}{(2\pi )^3}a^L_{n\ell}(\bold k)\,{\rm{exp}}(i \bold k \bold x-i Et),\nonumber \\
		a^L_{n\ell}(\bold k)&=&N^L k_{\perp}^{|\ell |}\,L_n^{|\ell |}(\frac{k_{\perp}^2}{\sigma _{\perp} ^2})\,\rm{exp}(-\frac{k_{\perp}^2}{2\sigma _{\perp} ^2}-\frac{(k_z-\bar k_z)^2}{2\sigma _z^2}+i \bold k \bold x-i Et+i \ell \varphi _k),
	\end{eqnarray}
	where $N^L$ is normalization coefficient, $k_{\perp}$ is the modulus of $\bold k$'s transverse part, $\sigma _{\perp}$ and $\sigma _z$ determine the transverse and longitudinal sizes of the state, $\bar k_z$ is the average longitudinal momentum, $L_n$ is the generalized Laguerre polynomial.
	
	For a fixed total final 4-momentum $(E_q,\,\bold q)$, the conservation of 3-momentum requires the three momenta $\bold q,\,\bold k_1,\,\bold k_2$ forming a triangle while the conservation of energy tells us $E_q=\sqrt{k_1^2+m_1^2}+\sqrt{k_2^2+m_2^2}$.
	These two conditions define a closed two-dimentional surface in momentum space.
	Specially, for zero mass case, it is an ellipsoid.
	A schematic analysis of interference space is displayed in Fig. \ref{an ellipsoid} for zero mass case.
	All the contribution points lie on an pink ellipsoid whose focal points happen to be endpoints of vector $\bold q$.
	$(\bold q-\bold k_2)$ and $\bold k_1$ distributions concentrate on two donuts.
	Main contribution to final state lies on intersection of the ellipsoid with the two donuts.
	
	The $\cal S$-matrix element for the process ``Two Laguerre-Gaussian vortex states to two plane wave states '' is
	\begin{eqnarray} \label{Laguerre-Gaussian collision}
		{\cal S}(L_1+L_2\rightarrow P_3+P_4)&\propto &\int d^3\bold k_1 d^3 \bold k_2 a^L_{n\ell _1}(\bold k_1) a^L_{n\ell _2}(\bold k_2){\cal M}\,\delta ^3(\bold k_1+\bold k_2-\bold q)\delta(E_1+E_2-E_q)\nonumber\\
		&\propto &\int _0^{\pi} d\hat \varphi_1\int _0^{\pi}d\hat \theta_1\,\frac{\sin \hat \theta_1 k_1^2E_1E_2\,g_1(k_{1\perp},k_{1z})g_2(k_{2\perp},k_{2z})}{|E_qk_1-E_1q\cos \hat \theta_1|}  \cos(\ell_1 \varphi_1+\ell_2 \varphi_2)\,{\cal M},
	\end{eqnarray}
	where
	\[ g_i(k_{i\perp},k_{iz})=k_{i\perp}^{|\ell _i|}\,L_{n_i}^{|\ell _i|}(\frac{k_{i\perp}^2}{\sigma _{i\perp} ^2})\,\rm{exp}(-\frac{k_{i\perp}^2}{2\sigma _{i\perp} ^2}-\frac{(k_{iz}-\bar k_{iz})^2}{2\sigma _{iz}^2}-i E_it),\,k_{i\perp}=k_i\sin \theta_i,\,k_{iz}=k_i\cos \theta_i,\,i=1,2.\]
	We also use two sets of coordinates systems which are the same with monochromatic vortex collision case. 
	In the second line of Eq. (\ref{Laguerre-Gaussian collision}), $(k_i,\varphi _i,\theta_i)$ are dependent on $(E_q,q,\theta _q,\hat \varphi_1,\hat \theta_1)$.
	At zero mass case, the relations are
	\begin{eqnarray}
		k_1&=&\frac{E_q^2-q^2}{2(E_q-q\, \cos \hat \theta_1)},\nonumber\\
		k_2&=&E_q-k_1,\nonumber\\
		\varphi_i&=&\arccos ({\frac{\cos \hat\theta _i\sin \theta _q+\sin \hat\theta_i\cos \theta_q \cos \hat\varphi _i}{\sqrt{1-(\cos \hat\theta _i \cos \theta _q-\sin \hat\theta _i\sin \theta _q\cos \hat\varphi _i)^2}}}),\nonumber\\
		\theta_i&=&\arccos (\cos \hat\theta_i\cos \theta_q-\sin \hat\theta_i\sin \theta_q\cos \hat \varphi_i)\nonumber
	\end{eqnarray}
	where $\theta_q$ is angle between vector $\bold q$ and $z$ axis and
	\begin{eqnarray}
		\hat \varphi_2&=&-\hat \varphi_1,\nonumber\\
		\hat \theta_2&=&\arccos(\frac{k_2^2+q^2-k_1^2}{2qk_2}).\nonumber
	\end{eqnarray}
	
	From Eq. (\ref{Laguerre-Gaussian collision}), we see that there  is no limitation on final total momentum or energy for collision of two Laguerre-Gaussian vortex states.
	Oscillation may appear in transverse distribution, longitudinal distribution and energy distribution with $E_q$, $q$ and $\theta _q$ varying.

	\section{results}	\label{section three}
	All the results shown below are based on the simple $\phi ^4$ scalar model with a constant amplitude ${\cal M}$ in tree level.
	For definiteness, we assume that the two colliding vortex states have the same momentum distributions except for different topological charges.
	
	\subsection{Bessel vortex collision}
	\begin{figure}[!h]
		\centering
		\includegraphics[width=0.45\textwidth]{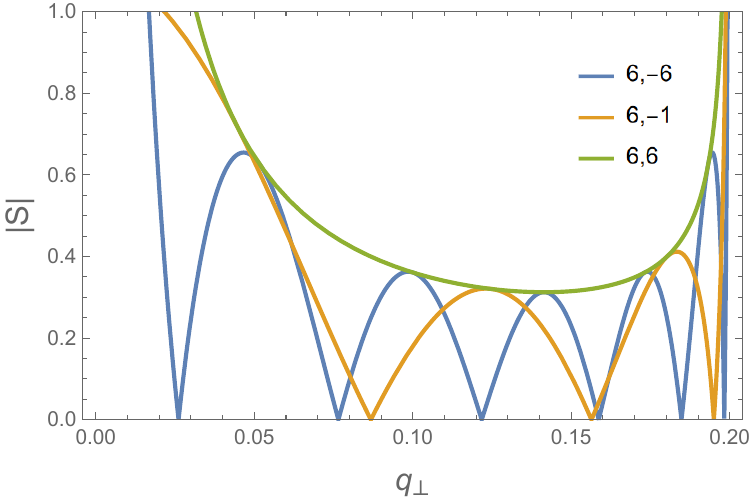}
		\includegraphics[width=0.45\textwidth]{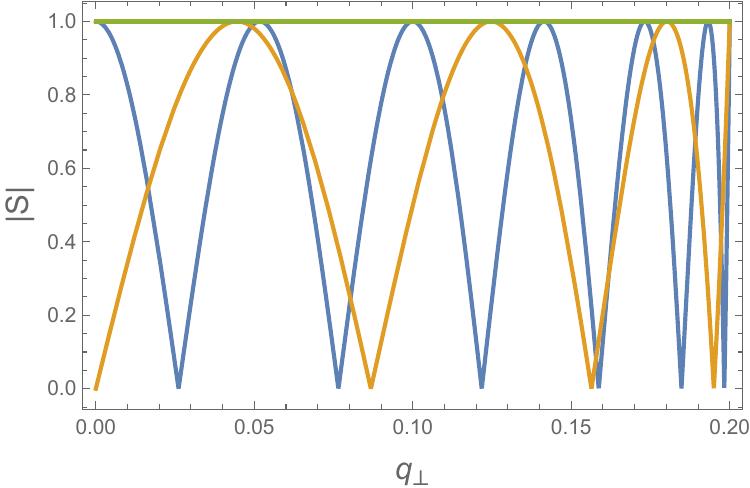}
		\caption{One-dimentional oscillation pattern of $|\cal S|$ distribution for Bessel vortex states collision. The distributions have been normalized so that biggest value of vertical coordinate is $1$. Left:  $|{\cal S}|$ oscillates when total final transverse momentum $q_{\perp}$ changes. Right: $\Delta \cdot |\cal S |$ oscillates when $q_{\perp}$ changes. The oscillation frequency is dependent on initial topological charges. Here three lines are corresponding to three cases that are different only by topological charges (Blue for $\ell_1=6,\,\ell_2=-6$; Brown for $\ell_1=6,\,\ell_2=-1$; Green for $\ell_1=6,\,\ell_2=6$). Other parameters of initial states: $E_1=E_2=2\,{\rm MeV},\,\kappa_1=\kappa_2=0.1\,{\rm MeV}$. Divergences have been cut out.}\label{Bessel oscillation}
	\end{figure}
	For the Bessel state collision, the total final energy and longitudinal momentum are fixed by the initial states, which is revealed by delta functions in Eq. (\ref{Bessel collision}).
	Thus, we only show radial transverse distributions within the range $|\kappa_1-\kappa_2|<q_{\perp}<|\kappa_1+\kappa_2|$, considering the distributions must be azimuthally symmetric.
	The value of $|{\cal S}|$ exhibits oscillations as a function of $q_\perp$ due to the factor $\cos (\ell _1 \varphi _1+\ell _2 \varphi _2)$.
	This oscillatory behavior can be seen as a momentum-space analogue of the famous two-slit interference pattern \cite{Elastic scattering of vortex electrons provides direct access to the Coulomb phase}.

	The value of $|{\cal S}|$ diverges near the boundaries of the $q_\perp$ region due to the denominator $\Delta$ in Eq.~\eqref{Bessel collision}. This unphysical behavior is the result of the initial Bessel states being non-normalizable in the transverse plane.
	Nevertheless, we still use Bessel states for comparison because they are a reasonable starting point for description of realistic vortex states, like the traditional plane waves are a good approximation of realistic wave packets.	
	
	Examples of the oscillation pattern are shown in Fig. \ref{Bessel oscillation}, and they confirm the above discussion.
	To demonstrate the oscillations more clearly, we removed in the right plot the divergence-inducing denominator $\Delta$.
	
	For the future discussion, it is convenient to pay attention to the points in momentum space where $|{\cal S}|$ passes through zero. 
	In the present example, we have several zero points, which break the entire $q_\perp$ domain into disjoint intervals.
	The number $n^B$ of these intervals are completely determined by the initial topological charges:
	\[n^B={\rm Integer}[|\ell_1-\ell_2|/2+1].\]
	Will this expression, which we derived for the Bessel states, remain the same if we use other forms of vortex states with the same $\ell$? The answer is negative, which we will verify in the following sections. 
	

	\subsection{General monochromatic vortex collision}
	\begin{figure}[!h]
		\centering
		\includegraphics[width=0.35\textwidth]{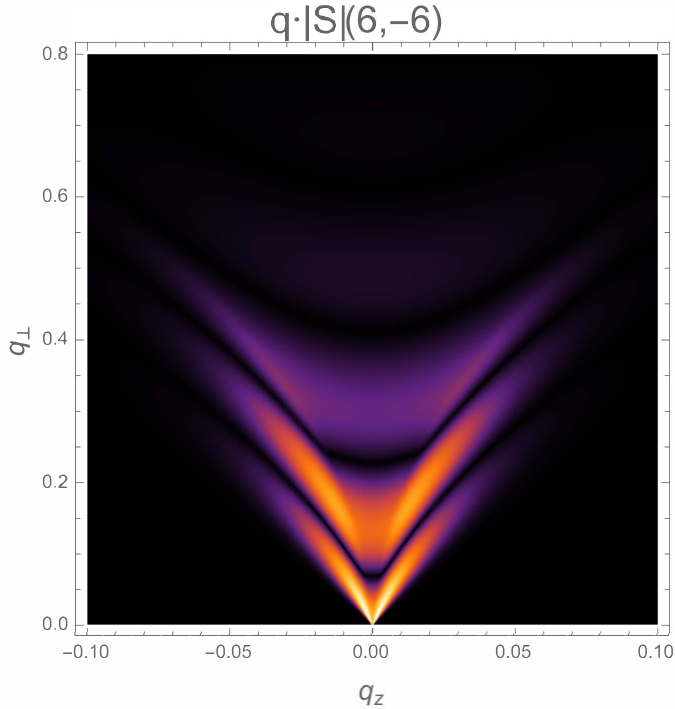}
		\includegraphics[width=0.06\textwidth,height=0.35\textwidth]{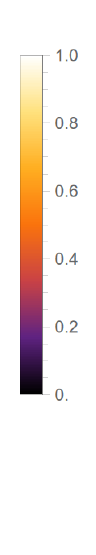}
		\includegraphics[width=0.35\textwidth]{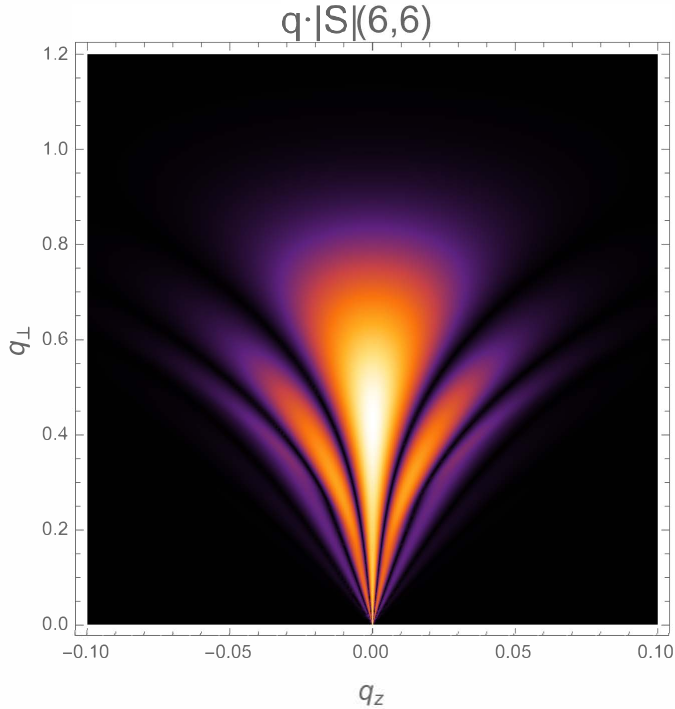}
		\includegraphics[width=0.06\textwidth,height=0.35\textwidth]{Legend}
		\\
		\includegraphics[width=0.35\textwidth]{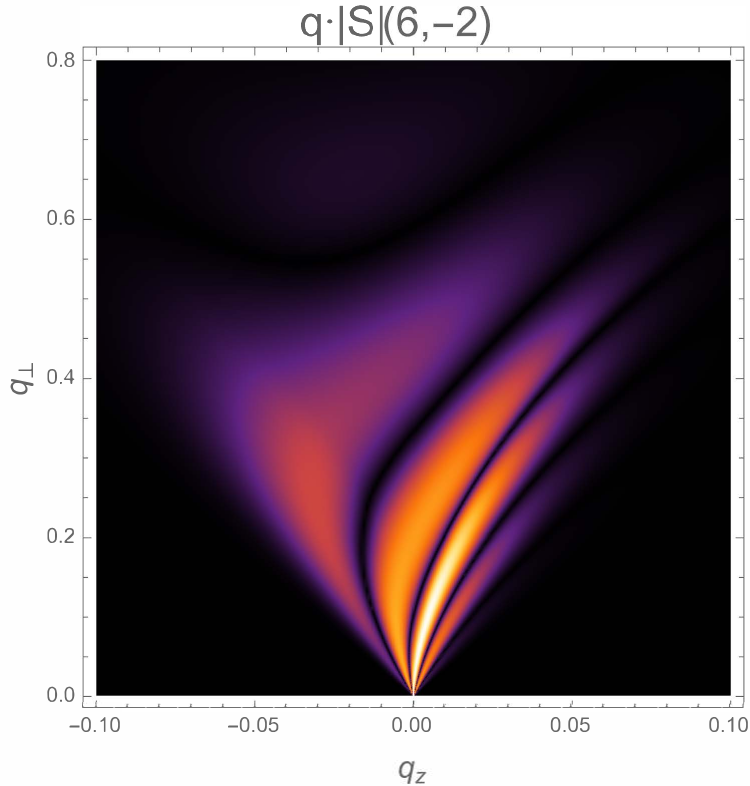}
		\includegraphics[width=0.06\textwidth,height=0.35\textwidth]{Legend}
		\includegraphics[width=0.35\textwidth]{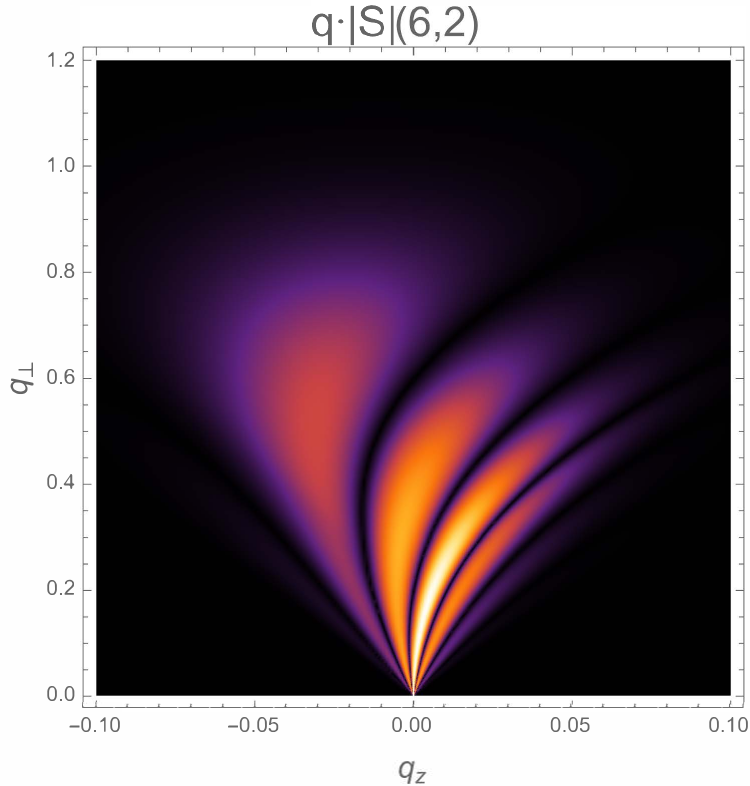}
		\includegraphics[width=0.06\textwidth,height=0.35\textwidth]{Legend}
		\caption{Two-dimentional oscillation pattern of $q\cdot|\cal S|$ distribution for general monochromatic vortex states collision. The distribution has been normalized so that its biggest value is $1$. We set $(\ell _1=-\ell _2=6)$ for left top  picture, $(\ell _1=\ell _2=6)$ for right top picture, $(\ell_1=6,\,\ell_2=-2)$ for left bottom picture and  $(\ell_1=6,\,\ell_2=2)$ for right bottom picture. Other parameters of initial states: $E_1=E_2=2 \,{\rm MeV},\,\theta_{01}=\theta_{02}=\sigma_{01}=\sigma_{02}=0.1\,{\rm rad}$. }\label{Monochromatic oscillation}
	\end{figure}
	\begin{figure}[t]
		\centering
		\includegraphics[width=0.35\textwidth]{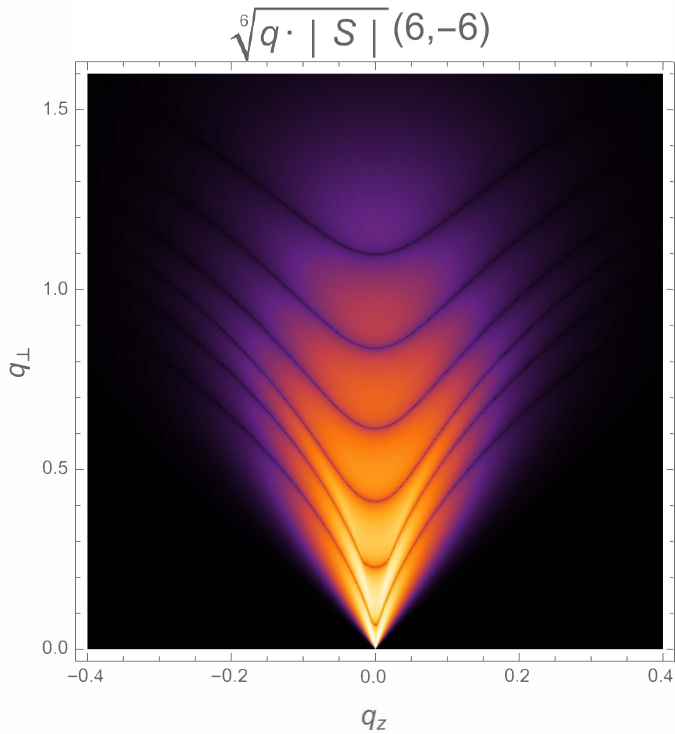}
		\includegraphics[width=0.06\textwidth,height=0.35\textwidth]{Legend}
		\includegraphics[width=0.35\textwidth]{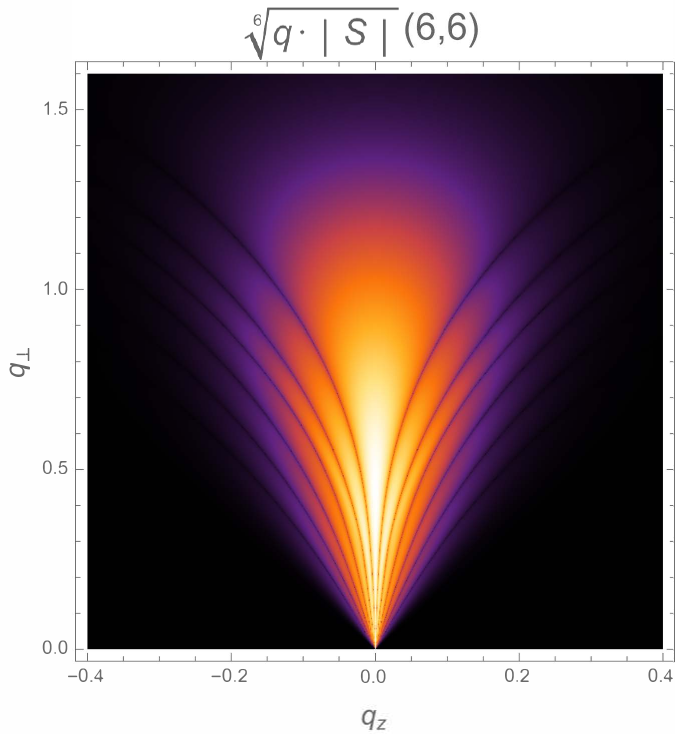}
		\includegraphics[width=0.06\textwidth,height=0.35\textwidth]{Legend}
		\\
		\includegraphics[width=0.35\textwidth]{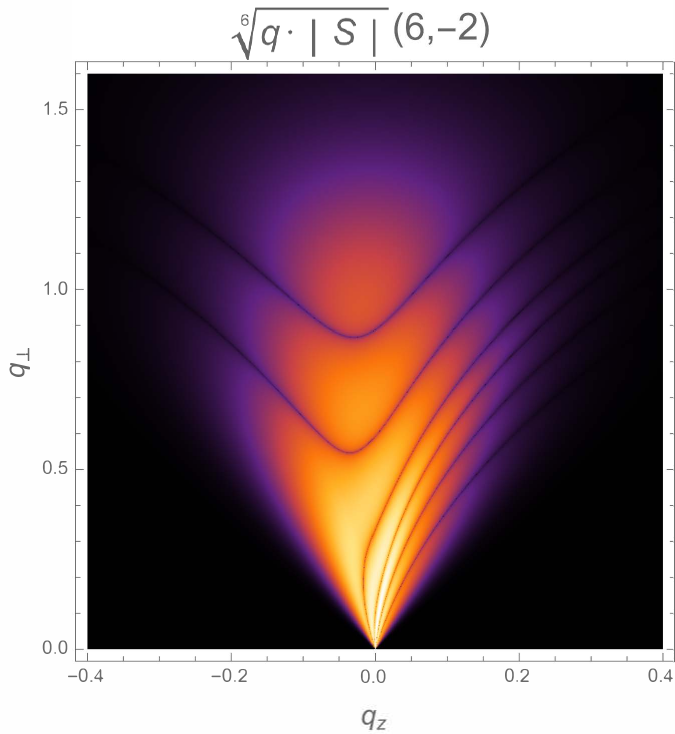}
		\includegraphics[width=0.06\textwidth,height=0.35\textwidth]{Legend}
		\includegraphics[width=0.35\textwidth]{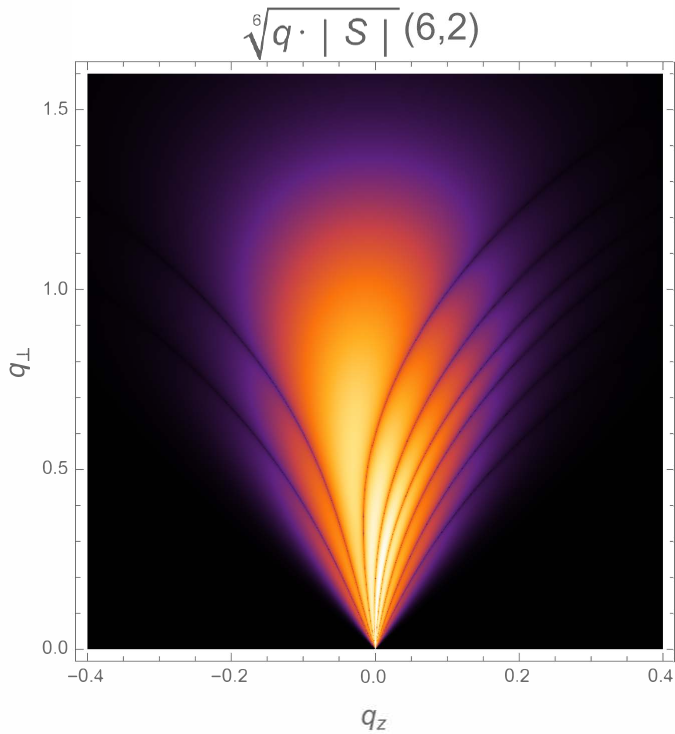}
		\includegraphics[width=0.06\textwidth,height=0.35\textwidth]{Legend}
		\caption{Two-dimentional oscillation patterns of $\sqrt[6]{q\cdot|\cal S|}$ distribution for general monochromatic vortex states collision. The distribution has been normalized so that its biggest value is $1$. We set $(\ell _1=-\ell _2=6)$ for left top  picture, $(\ell _1=\ell _2=6)$ for right top picture, $(\ell_1=8,\,\ell_2=-4)$ for left bottom picture and  $(\ell_1=8,\,\ell_2=4)$ for right bottom picture. Other parameters of initial states: $E_1=E_2=2 \,{\rm MeV},\,\theta_{01}=\theta_{02}=\sigma_{01}=\sigma_{02}=0.1\,{\rm rad}$. }\label{Monochromatic oscillation nth root 4}
	\end{figure}
	For general monochromatic vortex collision, only total final energy is fixed by initial states, as the delta function in Eq. (\ref{Bessel collision}) reveals.
	Thus, we will get both transverse and longitudinal distributions.
	The two-dimentional  distribution has oscillatory behavior which depends on topological  charges of initial states.
	Boundary of the distribution is determined by momentum modulus (i.e. radius of momentum distribution sphere) of the two states: $||\vec k_1|-|\vec k_2||<q<||\vec k_1|+|\vec k_2||$.
	The oscillation seems an analogue to the Fresnel diffraction in momentum space since the interference region is a circle as Fig. \ref{a circle} shows.
	But, at different point on the circle, weight factors are different.
	Thus it is actually an analogue to the case that the Fresnel diffraction experiment is prepared with an inhomogeneous wave source.
	
	As factor $1/q$ in Eq. \ref{Monochromatic collision} reveals, $|\cal S|$ distribution is highly concentrated near point $q=0$.
	The asymptotic behavior at this value is:
	\[\lim_{q\rightarrow 0}|{\cal S}|=\begin{cases}
		\infty   & \mathrm{if,} \ \ell_1+\ell_2\equiv 0\,;\\
		0  & \mathrm{if,} \ \ell_1+\ell_2\neq 0\,.
	\end{cases}\]
	The divergence is a direct result of the non-normalizable initial states.
	They can be normalized in transverse plane but can not be normalized in longitudinal direction.
	This means the state can be seen as a beam state but not a particle state.
	To see clearly the oscillatory behavior, we will show $q\cdot|\cal S|$ instead of $|\cal S|$ distribution.
	It rearranges the distribution but will not influence oscillatory behavior.
	The results are shown in Fig. \ref{Monochromatic oscillation}.
	Two-dimentional oscillation patterns show that the distribution not only oscillates with transverse momentum changing but also with longitudinal momentum changing.
	This is a new feature that is completely different with Bessel vortex case.
	This appears because the interference space (the circle shown in Fig. \ref{a circle}) includes points with both different transverse momentum and different longitudinal momentum.
	
	Another important result is that the oscillation pattern is partly smeared out after integration in Eq. (\ref{Monochromatic collision}).
	There is obvious tendency that value of bright oscillation peaks decreases and some peaks may just run out of our sight with much small amplitudes.
	A more careful treatment do find more peaks and will be shown later.
	The smeared-out result is analogous to the smeared-out pattern in multi-frequency double-slit interference \cite{Quantum interference with molecules: The role of internal states}.
	Due to energy dispersion of initial states, some of bright peaks in final interference pattern significantly decreases and seems disappear.
	In fact, the interference space shown in Fig. \ref{a circle} can be seen as combination of many ``two-slits'' which are characterized by different $\theta$.
	By weight function Eq. (\ref{weight fucntion}), all the ``two-slits'' are assigned with different weight factors and the final interference pattern are partly smeared out.
	
	There is an interesting result that the interference pattern is symmetric under reverse of longitudinal momentum (or forward-backward symmetric) if topological charges of initial states have same absolute values.
	Or else, it will be forward-backward asymmetric.
	This is different with Bessel vortex collision case in which asymmetric information that is only induced by two topological charges can not be reserved in final distribution.
	However, the asymmetric information that is only induced by sign of topological charges can not be kept.
	This is not difficult to be understood.
	Three kinds of transformations should be considered here: parity transformation in $z$ direction, reverse of topological charges and parity transformation in transverse plane along any direction.
	For initial states which satisfy $|\ell_1|=|\ell_2|$, The first transformation is equivalent to the second one which is equivalent to the third one.
	Since we are only interested with rotational-symmetric total final momentum distribution which do not change under reverse of transverse momentum in any direction, it should also be symmetric when we reverse longitudinal momentum.
	While for the case $|\ell_1|\neq|\ell_2|$, the first transformation is not equivalent to the second one.
	
	The most attractive feature for collision of two general monochromatic vortex states lies in distributions that are reshaped by $n$th root, which are shown in Fig. \ref{Monochromatic oscillation nth root 4}.
	Only four examples are shown in this figure and more examples will be listed in Appendix A Fig. \ref{Monochromatic oscillation nth root}.
	All the $q\cdot|\cal S|$ values are redefined by the ``$n$th root'', except for zero ones.
	Thus we can see all the unconnected areas, which are partitioned by zero lines, in the distribution and find special topological structures that are completely dependent on initial topological charges.
	Number of the unconnected areas can be expressed as function of initial topological charges:
	\begin{eqnarray}
		&&n^{M}=n_{\perp}^M+n_{z}^M-1,\,\,\,\,\,\,n_{\perp}^M=\frac{|\ell_1|+|\ell_2|-|\ell_1+\ell_2|}{2}+1,\,\,\,\,\,\,
		n_{z}^M=|\ell_1+\ell_2|+1.\nonumber
	\end{eqnarray}
	 The superscript ``$M$'' represents initial general monochromatic vortex states.
	 $n^M$ is total number.
	According to transverse and longitudinal oscillation, another two kinds of numbers $n_{\perp}^M$ and $n_z^M$ are defined.
	They are corresponding to count area number in transverse and longitudinal direction respectively.
	The decompose of topological numbers seems nonsense now.
	Meanwhile, it will make sense after comparison with Laguerre-Gaussian vortex collision case.

	\subsection{Laguerre-Gaussian vortex collision}
	\begin{figure}[t]
		\centering
		\includegraphics[width=0.35\textwidth]{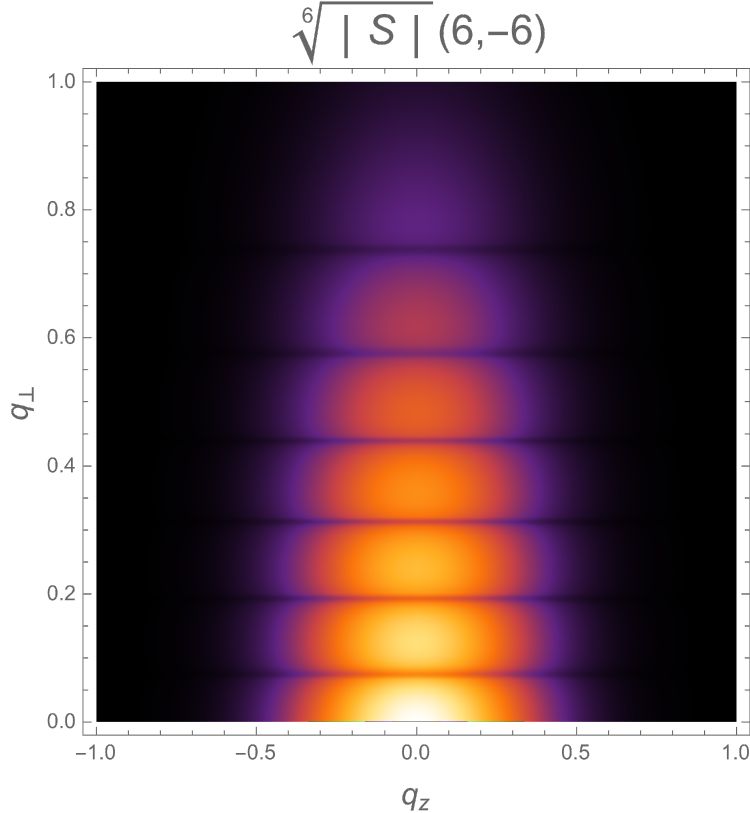}
		\includegraphics[width=0.06\textwidth,height=0.35\textwidth]{Legend}
		\includegraphics[width=0.35\textwidth]{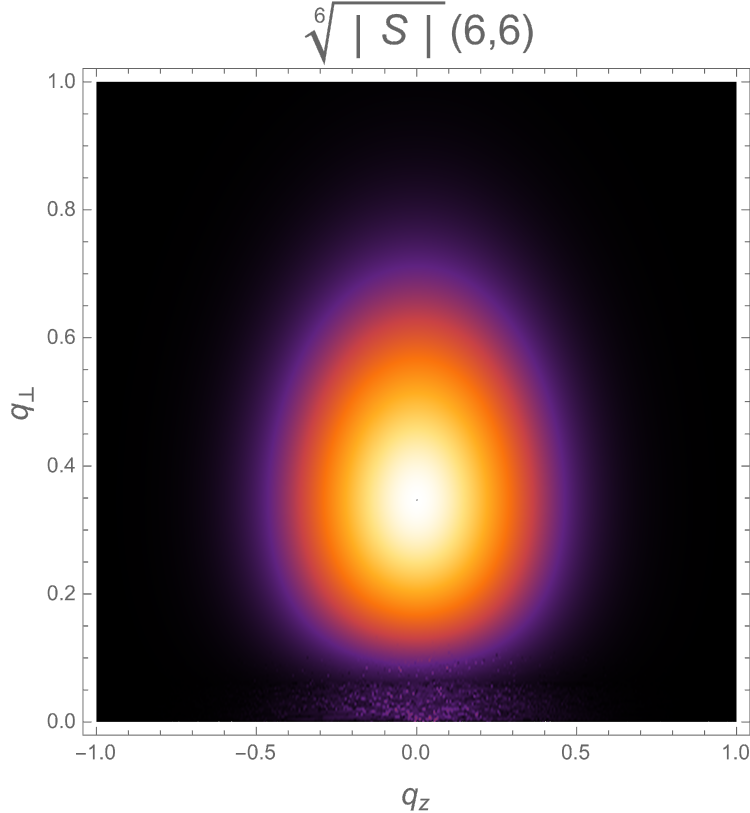}
		\includegraphics[width=0.06\textwidth,height=0.35\textwidth]{Legend}
		\\
		\includegraphics[width=0.35\textwidth]{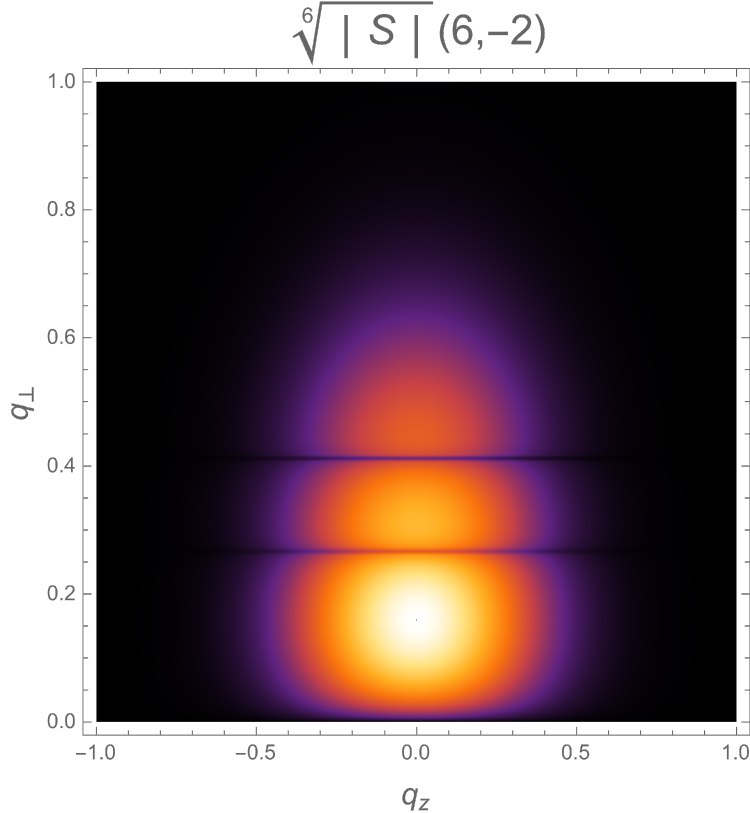}
		\includegraphics[width=0.06\textwidth,height=0.35\textwidth]{Legend}
		\includegraphics[width=0.35\textwidth]{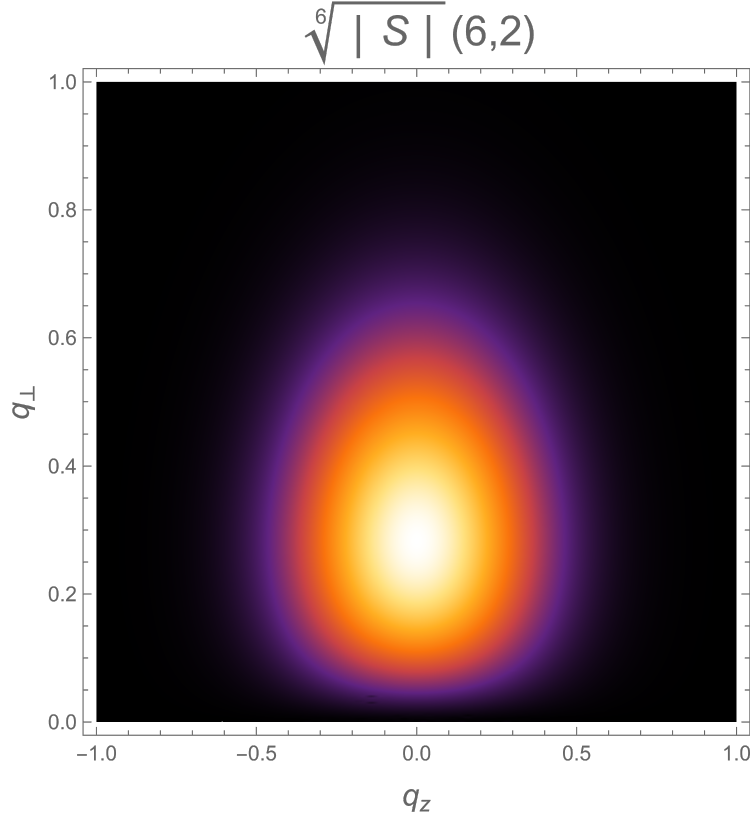}
		\includegraphics[width=0.06\textwidth,height=0.35\textwidth]{Legend}
		\caption{Two-dimentional oscillation patterns of  $\sqrt[6]{|\cal S|}$ distribution at fixed total energy ($E_q=4\,{\rm MeV}$) for Laguerre-Gaussian vortex states collision. The distribution has been normalized so that its biggest value is $1$. We set $(\ell _1=-\ell _2=6)$ for left top  picture, $(\ell _1=\ell _2=6)$ for right top picture, $(\ell_1=6,\,\ell_2=-2)$ for left bottom picture and  $(\ell_1=6,\,\ell_2=2)$ for right bottom picture.  Other parameters of initial states: $\bar k_{z1}=\bar k_{z2}=2 \,{\rm MeV},\,\sigma_{z1}=\sigma_{z2}=\sigma_{\perp 1}=\sigma_{\perp 2}=0.1\,{\rm MeV},\,n_1=n_2=0$.}\label{LG oscillation nth root 4}
	\end{figure}
	For Laguerre-Gaussian vortex collision, we can not only get total momentum distribution in transverse and longitudinal direction, we can also get total energy distribution of final plane waves.
	There is no certain boundary for the distribution since momentum components of  initial states run out of all the momentum space, though some of them are  more probable than others.
	The interference region is an ellipsoid and there is no analogue to this kind of interference.
	
	Two dimentional distributions in $xOy$ plane with fixed total energy are shown in Fig. \ref{LG oscillation nth root 4}.
	Only four examples are shown and more examples will be listed in Appendix A Fig. \ref{LG oscillation}.
	As numerical results as well as Fig. \ref{LG oscillation} reveal, oscillation only appears in radial direction of transverse plane though total final momentum distribution has three dimentions.
	This can only be explained by effect of smearing due to the two-dimentional integral in Eq. (\ref{Laguerre-Gaussian collision}).
	We have seen that the distribution of general monochromatic vortex collision been partly smeared out by one-dimentional integral.
	For Laguerre-Gaussian case, two-dimentional integral works more effectively.
	Oscillation of final distribution in longitudinal and energetic direction are completely smeared out, while oscillation in transverse direction is just partly smeared out.
	An direct result of this changing is that the topological charge-induced forward-backward asymmetry is difficult to be seen in Laguerre-Gaussian case.
	The distributions also display topological structures in this case.
	According to pictures with different initial topological charges, we conclude that the topological charge dependence is:
	\begin{eqnarray}
		&&n^{L}=n_{\perp}^L+n_{z}^L-1,\,\,\,\,\,\,n_{\perp}^L=\frac{|\ell_1|+|\ell_2|-|\ell_1+\ell_2|}{2}+1,\,\,\,\,\,\,n_{z}^L=1.\nonumber
	\end{eqnarray}
	The superscript ``$L$'' represents initial Laguerre-Gaussian vortex states.
	We can see that $n_{\perp}^L$ and $n_{\perp}^M$ have same topological charge dependence.
	This is a nice result and means that it may be a general principle for physical vortex states collision.
	Thus it may be used for measurement of vortex state's topological charge.
	
\begin{figure}[!h]
		\centering
		\includegraphics[width=0.45\textwidth]{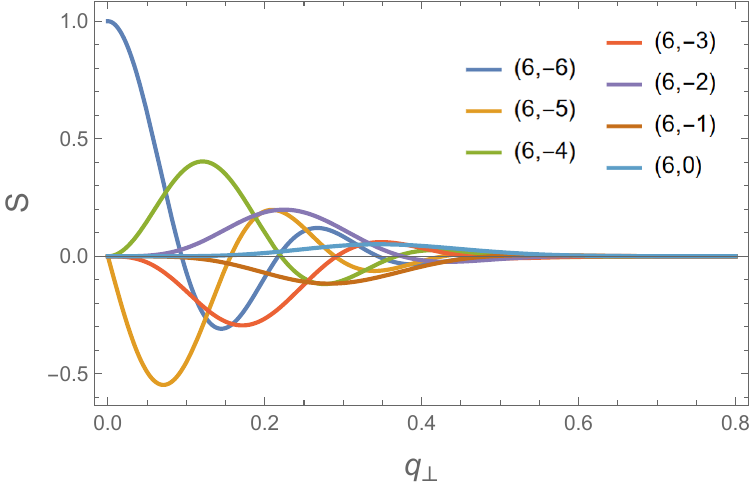}
		\includegraphics[width=0.45\textwidth]{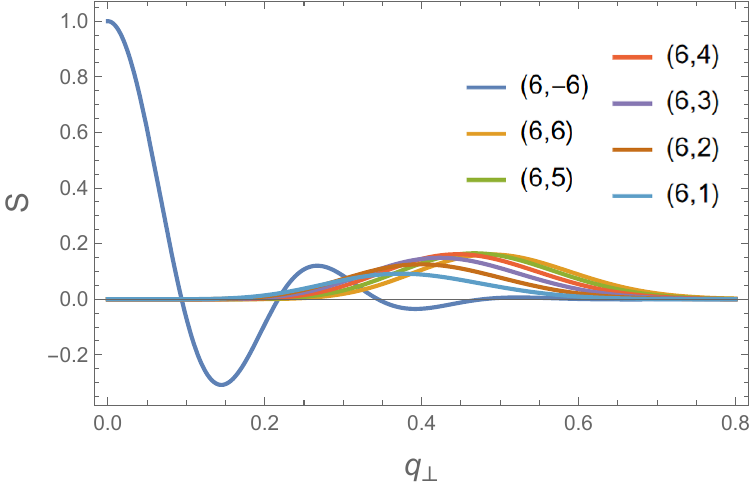}
		\caption{One-dimentional $\cal S$ distributions at fixed total energy ($E_q=4\,{\rm MeV}$) and total longitudinal momentum ($q_z=0\,{\rm MeV}$) for Laguerre-Gaussian vortex states collision. The distributions have been normalized so that biggest value of vertical coordinate is $1$. Left: $\ell_1=6$ and $\ell_2=-6,\,-5,\,-4,\,-3,\,-2,\,-1,\,0$. Right: $\ell_1=6$ and $\ell_2=-6,\,1,\,2,\,3,\,4,\,5,\,6$. Other parameters of initial states: $\bar k_{z1}=\bar k_{z2}=2 \,{\rm MeV},\,\sigma_{z1}=\sigma_{z2}=\sigma_{\perp 1}=\sigma_{\perp 2}=0.1\,{\rm MeV},\,n_1=n_2=0$. }\label{LG oscillation 1D}
	\end{figure}

	Another obvious feature is that the distribution near $z$ axis tends to zero for almost all sub-pictures in Fig. \ref{LG oscillation} except for the case that satisfies $\ell_1+\ell_2\equiv 0$.
	We see that distribution near $z$ axis is non-zero and relatively large for this special case.
	In fact, this effect can be calculated analytically by Eq. (\ref{Laguerre-Gaussian collision}).
	Setting $q_z$ (longitudinal part of total momentum $\bold q$) equal to zero, we find that
	\begin{eqnarray}\label{LG collision 1D}
		{\cal S}(L_1+L_2\rightarrow P_3+P_4)\propto \int _0^{\pi} d \varphi_1 \,\cos[(\ell_1+\ell_2) \varphi_1)]\,{\cal M},
	\end{eqnarray}
	where another integral in Eq. (\ref{Laguerre-Gaussian collision}) is not shown here because the two integral are completely separated by setting $q_{z}=0$.
	Eq. (\ref{LG collision 1D}) is non-zero only if $\ell_1+\ell_2=0$.
	This feature characterizes the initial system of two vortex states with zero total angular momentum.
	It can also be used for measurement of vortex particle's topological charge.
	By the way, oscillation of Laguerre-Gaussian vortex collision can also been shown by one-dimensional transverse distribution since longitudinal and energetic distributions are completely smeared out.
	Pictures of this kind are shown in Fig. \ref{LG oscillation 1D}.
	By these figures, the feature is much more clear.
	
	\section{Conclusions and the outlook}\label{section four}
	In this work, using qualitative calculations and discussions, 
	we investigated the peculiar topological features which arise in the final momentum distributions
	in scalar vortex states collisions.
	We compared the results for three kinds of vortex states: the Bessel, a general monochromatic, and the Laguerre-Gaussian vortex states. The main findings are as follows.
	\begin{itemize}
		\item 
		The total final momentum distribution possesses peculiar topological structure in momentum space in the following sense: the distribution is partitioned into several concentric doughnut-like regions by multiple ${\cal S} = 0$ surfaces.
		The number of these disjoint regions is fully determined by the topological charges of the initial states.
		\item 
		The total final momentum distributions show an additional feature which is characteristic for the case $\ell_1+\ell_2=0$,
		which means that the angular momentum of the entire initial states is zero.
		Namely, only in this case can the distribution in the vicinity of the $z$ axis be non-zero and relatively large.
		This is especially clear for the Laguerre-Gaussian vortex collision case.
		Note that most experimental vortex states are formulated as Laguerre-Gaussian states.
	\end{itemize}
	Both features provide methods to measure an unknown topological charge of a vortex particle (or beam) when it collides 
	with a vortex target with a known topological charge.
	Whether they can be applied to realistic scattering of vortex states requires detailed quantitative computations.
	These calculations should be done with the Laguerre-Gaussian vortex particles because they are readily generated in experiments. 
	
	%
	%
	%
	%
	
	
	Other properties of vortex particle collision discussed in this paper include:
	a ``smearing-out'' effect, analogous to what is seen in two-slit interference 
	and an unusual forward-backward asymmetry induced by $|\ell_1| \not = |\ell_2|$, 
	which manifests itself only for the general monochromatic vortex particles.
	We hope that all these features will remain in realistic collisions of vortex electrons and other particles.
	
	\section*{Acknowledgments}
	
	We thank professor Igor P. Ivanov in School of Physics and Astronomy in Sun Yat-sen University for instructive suggestions and helpful discussions.
	This work was supported by grants of the National Natural
	Science Foundation of China (Grant No. 11975320) and the
	Fundamental Research Funds for the Central Universities,
	Sun Yat-sen University.
	
	\appendix 
	\section{Dependence of the momentum space distributions on $\ell$}
	Here, for completeness, we show how the momentum space distributions for the monochromatic and Laguerre-Gaussian states 
	depend on the values of the topological charge $\ell_2$ with a fixed $\ell_1$.

	\begin{figure}[!h]
		\centering
		\includegraphics[width=0.25\textwidth]{M_6--6_wide}
		\includegraphics[width=0.038\textwidth]{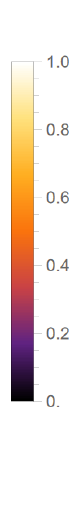}
		\includegraphics[width=0.25\textwidth]{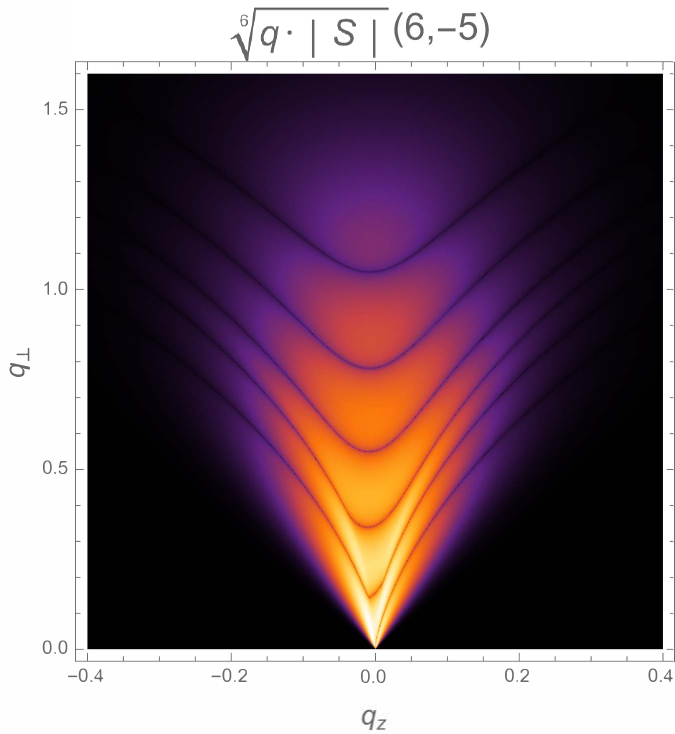}
		\includegraphics[width=0.038\textwidth]{Legend_2}
		\includegraphics[width=0.25\textwidth]{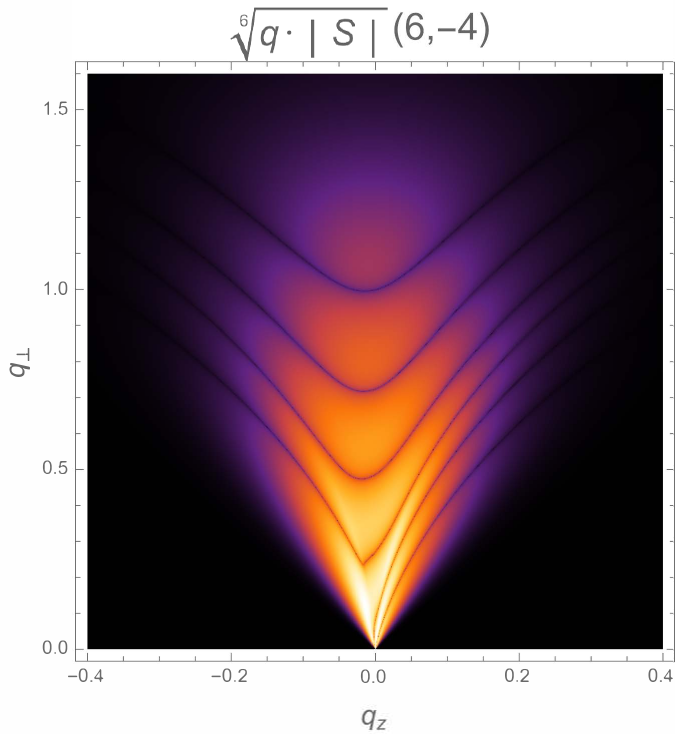}
		\includegraphics[width=0.038\textwidth]{Legend_2}
		\\
		\includegraphics[width=0.25\textwidth]{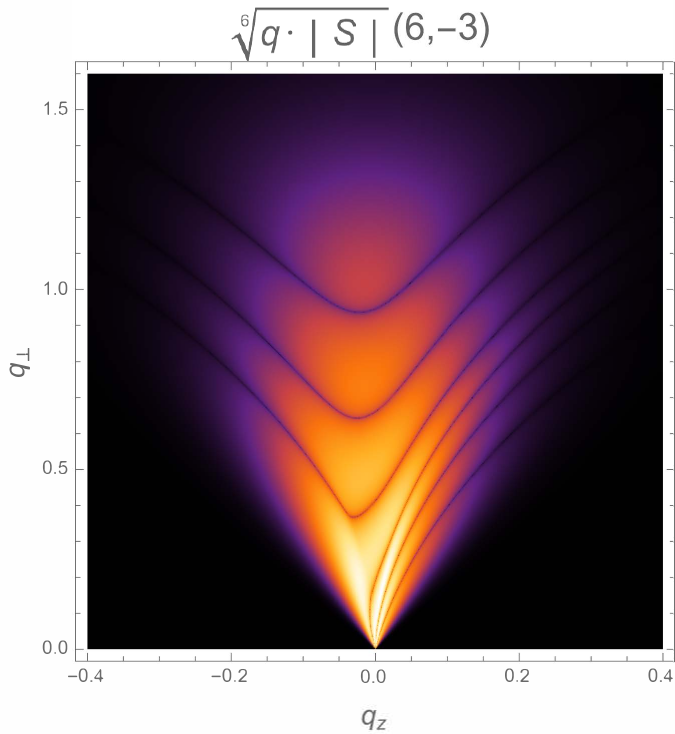}
		\includegraphics[width=0.038\textwidth]{Legend_2}
		\includegraphics[width=0.25\textwidth]{M_6--2_wide}
		\includegraphics[width=0.038\textwidth]{Legend_2}
		\includegraphics[width=0.25\textwidth]{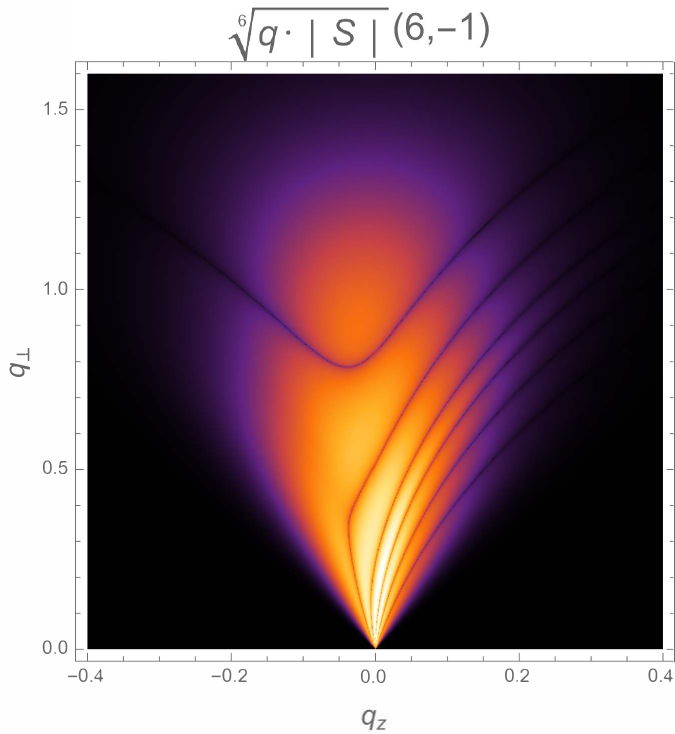}
		\includegraphics[width=0.038\textwidth]{Legend_2}
		\\
		\includegraphics[width=0.25\textwidth]{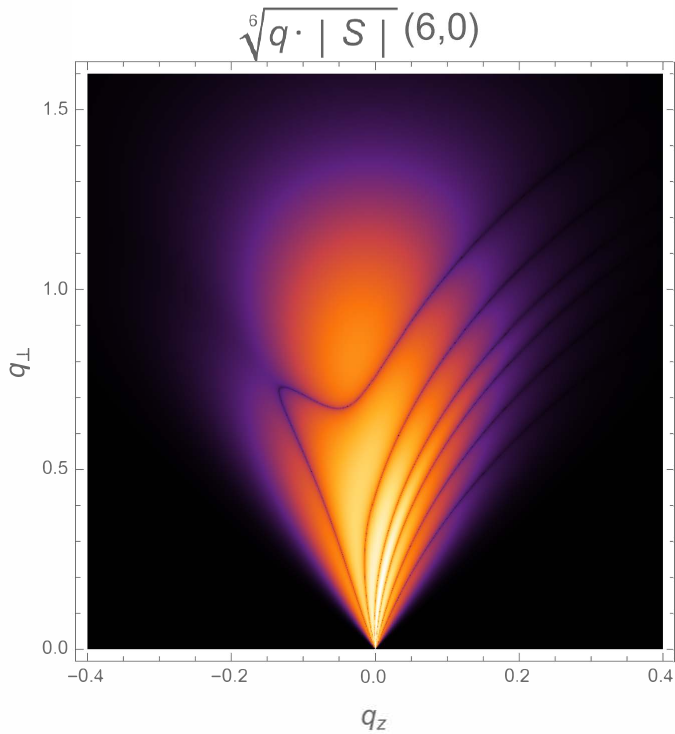}
		\includegraphics[width=0.038\textwidth]{Legend_2}
		\includegraphics[width=0.25\textwidth]{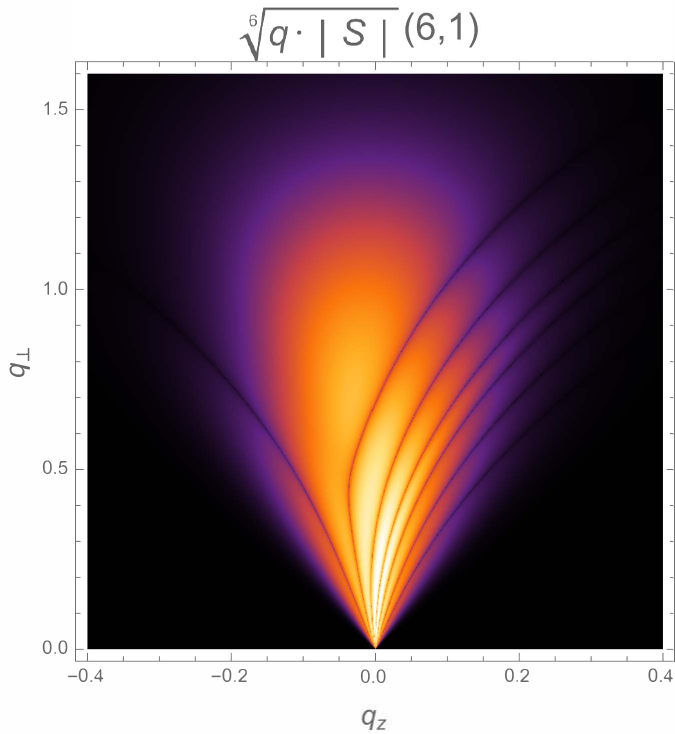}
		\includegraphics[width=0.038\textwidth]{Legend_2}
		\includegraphics[width=0.25\textwidth]{M_6-2_wide}
		\includegraphics[width=0.038\textwidth]{Legend_2}
		\\
		\includegraphics[width=0.25\textwidth]{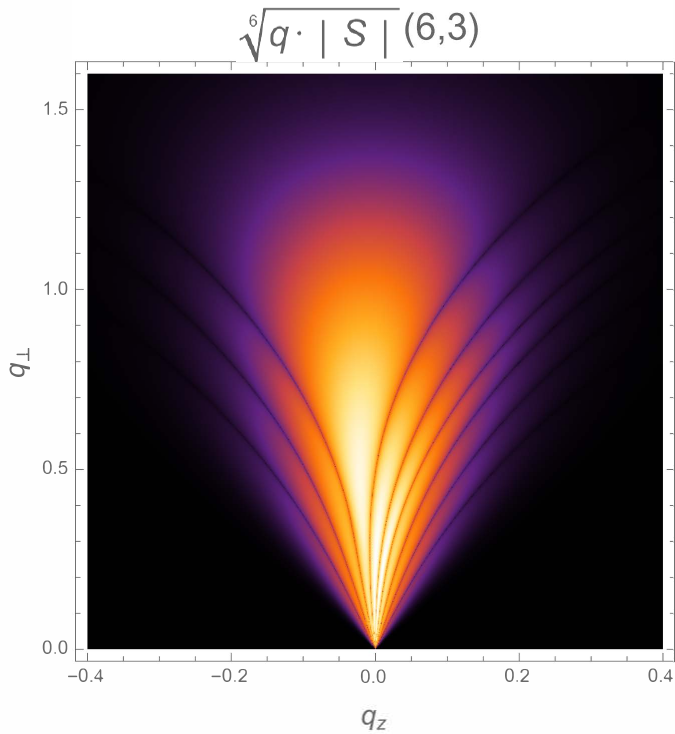}
		\includegraphics[width=0.038\textwidth]{Legend_2}
		\includegraphics[width=0.25\textwidth]{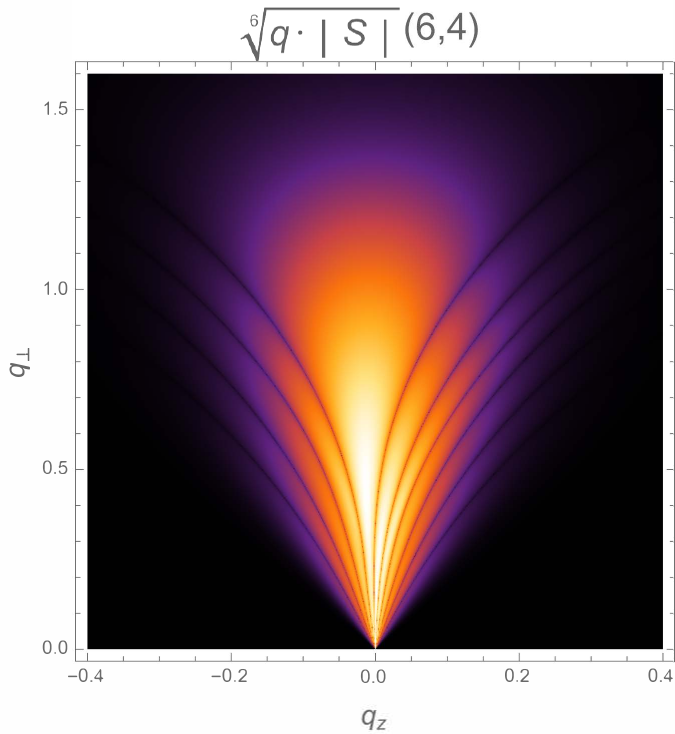}
		\includegraphics[width=0.038\textwidth]{Legend_2}
		\includegraphics[width=0.25\textwidth]{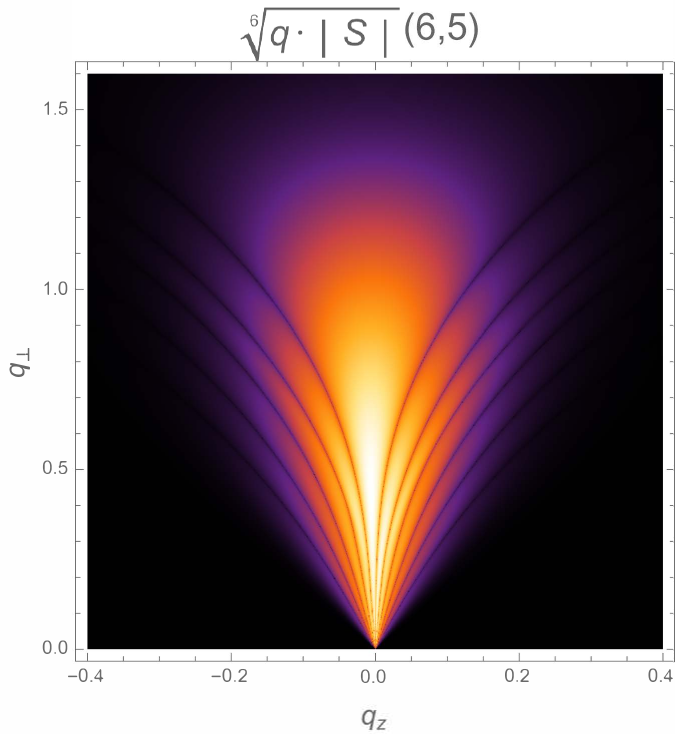}
		\includegraphics[width=0.038\textwidth]{Legend_2}
		\caption{Two-dimentional oscillation patterns of $\sqrt[6]{|\cal S|}$  distribution for general monochromatic vortex states collision. The distribution has been normalized so that its biggest value is $1$. $(\ell_1,\,\ell_2)$ is shown in every sub-picture. Other parameters of initial states: $E_1=E_2=2 \,{\rm MeV},\,\theta_{01}=\theta_{02}=\sigma_{01}=\sigma_{02}=0.1\,{\rm rad}$.}\label{Monochromatic oscillation nth root}
	\end{figure}\FloatBarrier

	\begin{figure}[!h]
		\centering
		\includegraphics[width=0.25\textwidth]{LG_6--6_wide}
		\includegraphics[width=0.038\textwidth]{Legend_2}
		\includegraphics[width=0.25\textwidth]{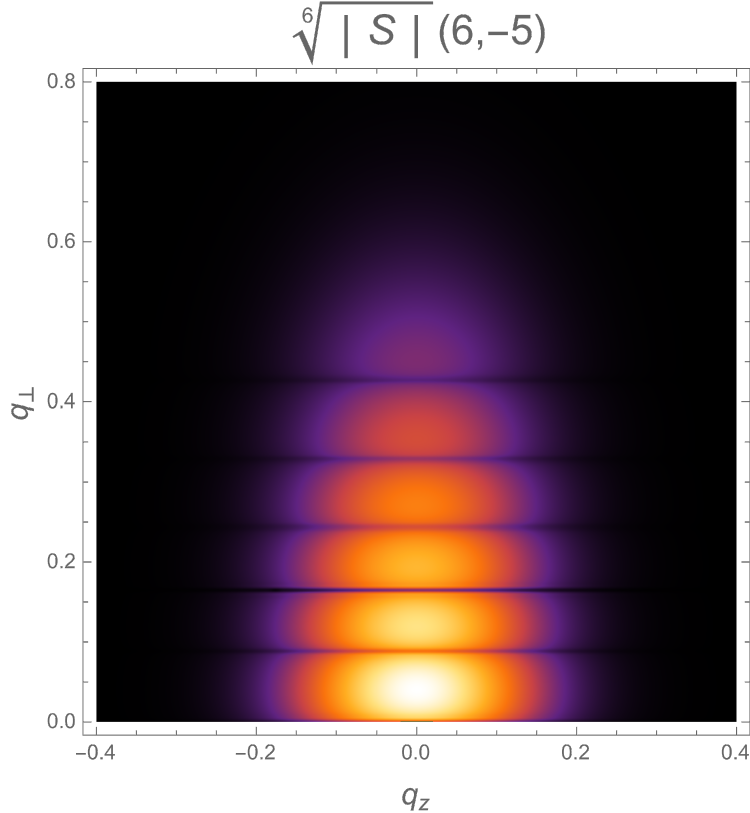}
		\includegraphics[width=0.038\textwidth]{Legend_2}
		\includegraphics[width=0.25\textwidth]{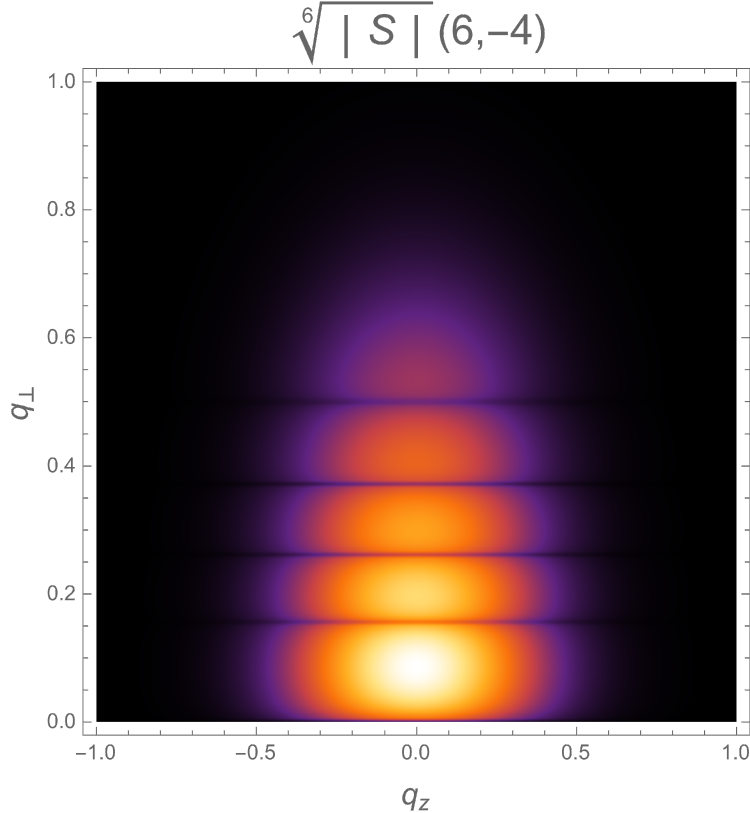}
		\includegraphics[width=0.038\textwidth]{Legend_2}
		\\
		\includegraphics[width=0.25\textwidth]{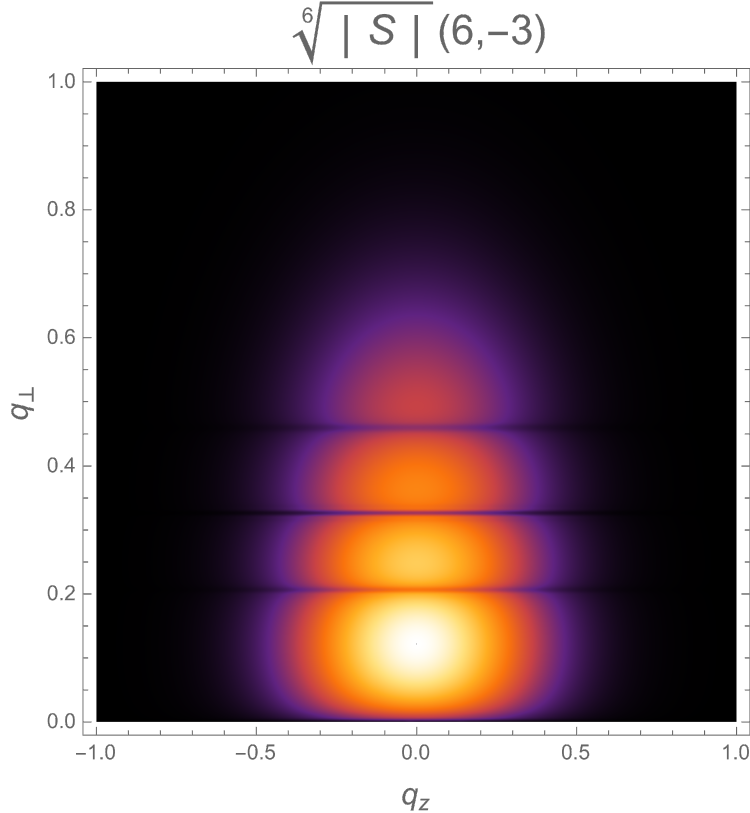}
		\includegraphics[width=0.038\textwidth]{Legend_2}
		\includegraphics[width=0.25\textwidth]{LG_6--2_wide}
		\includegraphics[width=0.038\textwidth]{Legend_2}
		\includegraphics[width=0.25\textwidth]{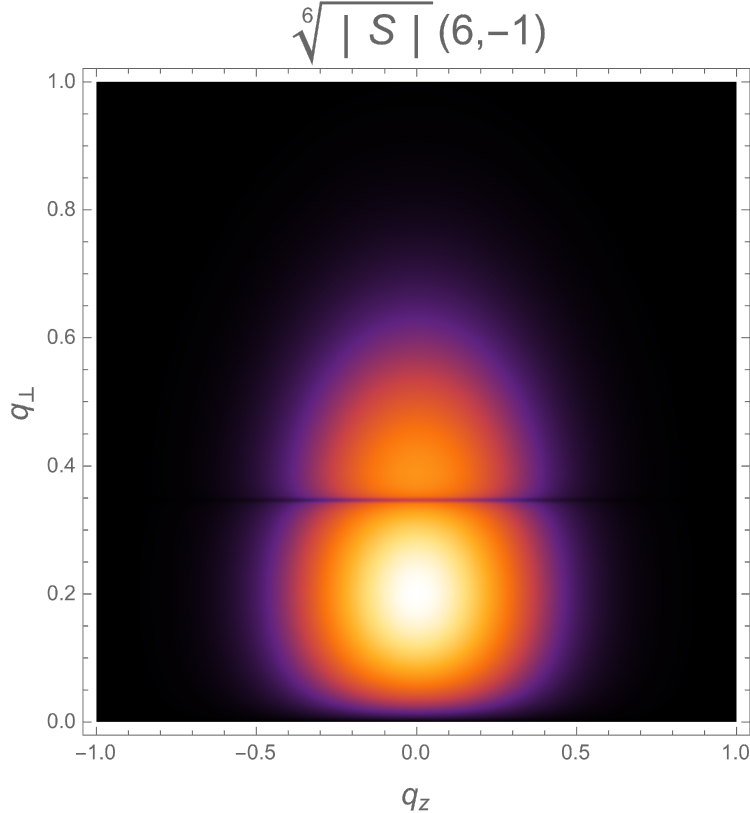}
		\includegraphics[width=0.038\textwidth]{Legend_2}
		\\
		\includegraphics[width=0.25\textwidth]{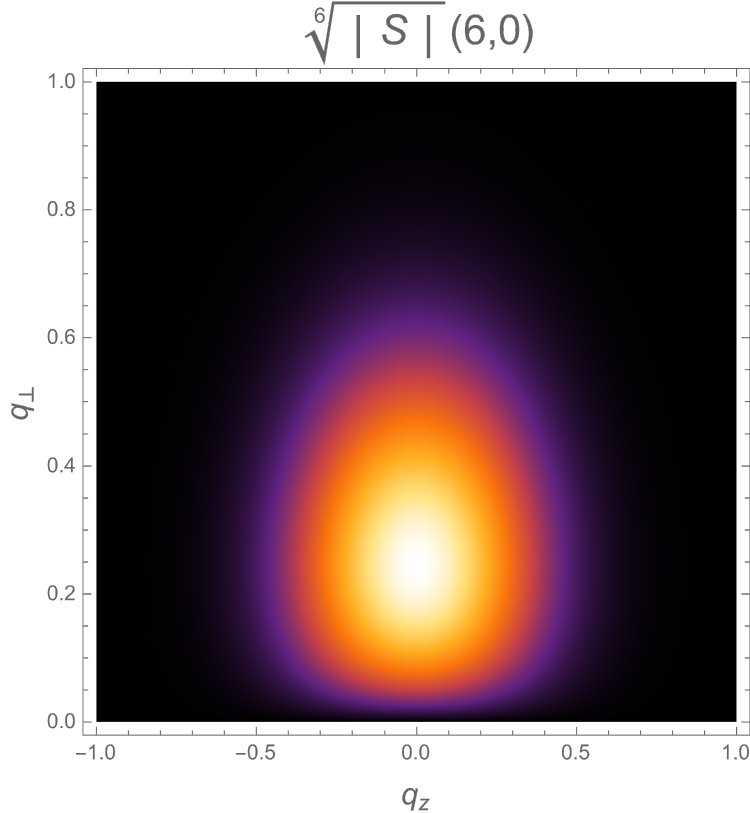}
		\includegraphics[width=0.038\textwidth]{Legend_2}
		\includegraphics[width=0.25\textwidth]{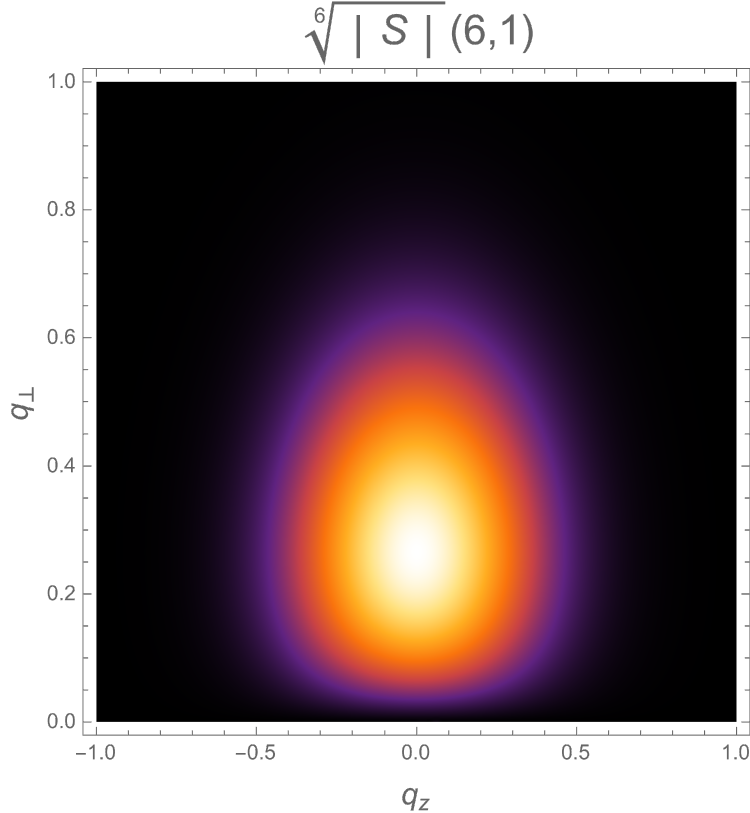}
		\includegraphics[width=0.038\textwidth]{Legend_2}
		\includegraphics[width=0.25\textwidth]{LG_6-2_wide}
		\includegraphics[width=0.038\textwidth]{Legend_2}
		\\
		\includegraphics[width=0.25\textwidth]{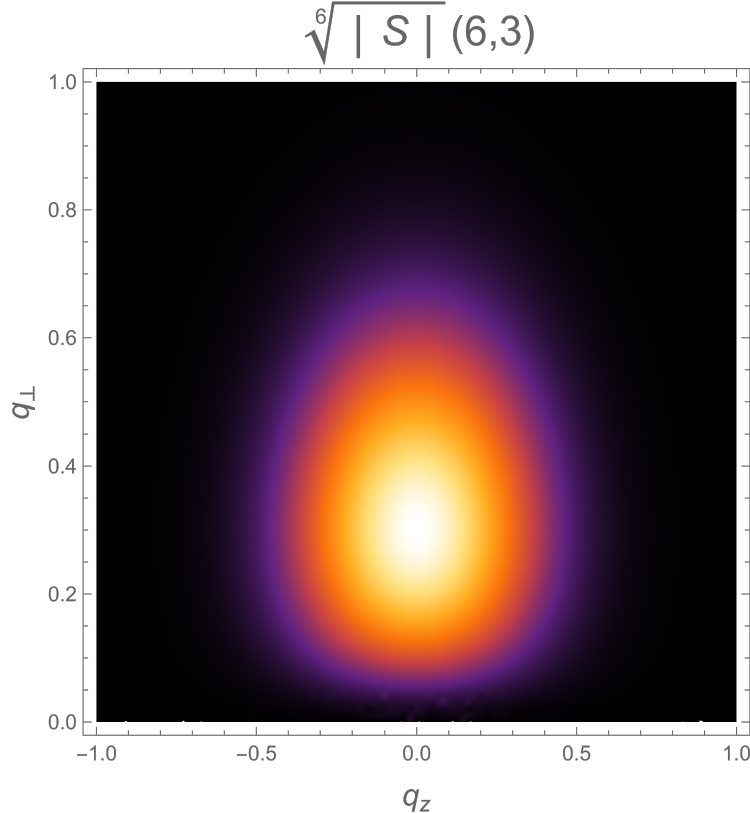}
		\includegraphics[width=0.038\textwidth]{Legend_2}
		\includegraphics[width=0.25\textwidth]{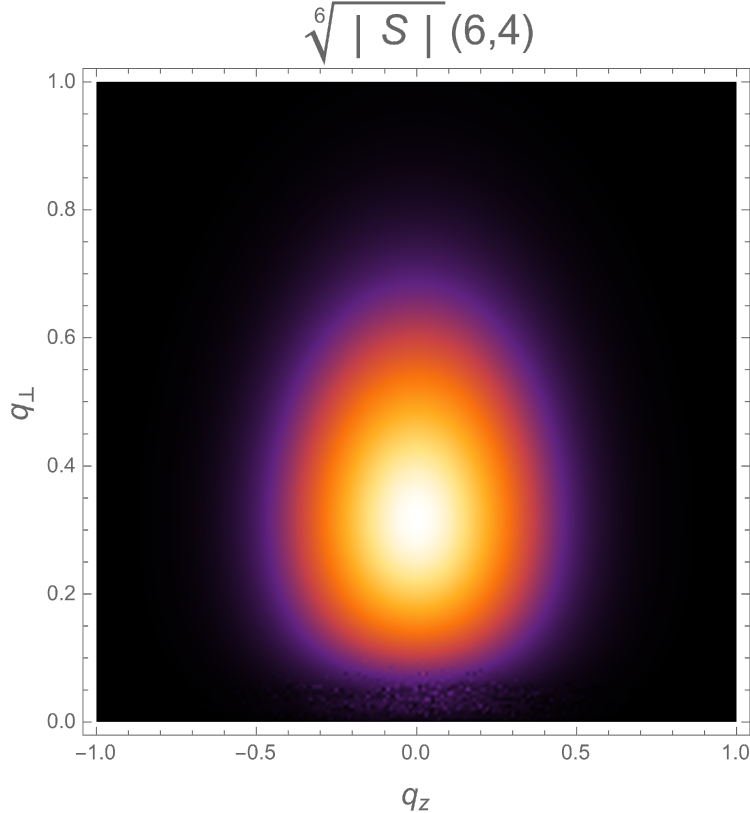}
		\includegraphics[width=0.038\textwidth]{Legend_2}
		\includegraphics[width=0.25\textwidth]{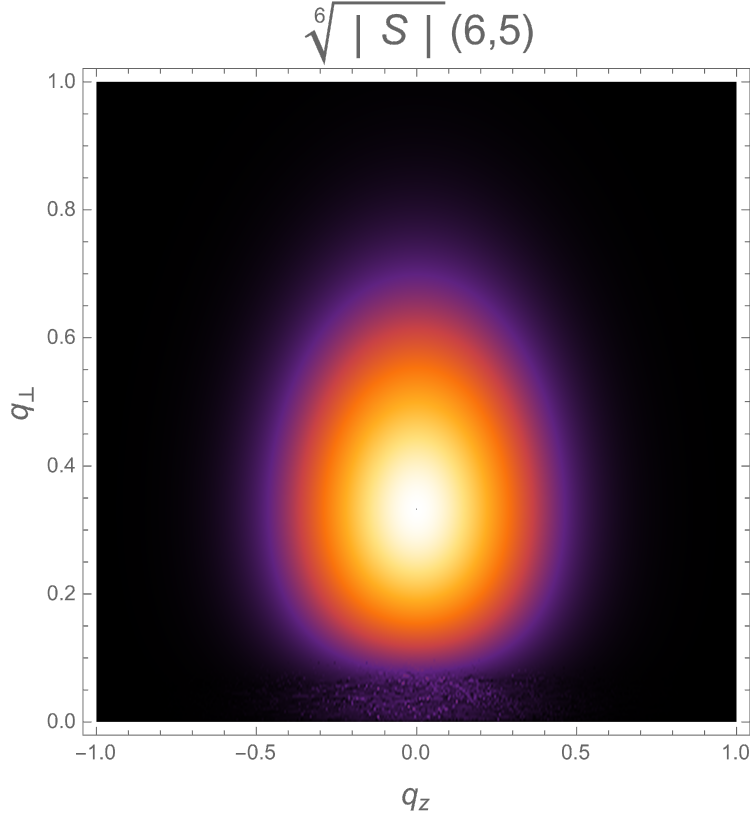}
		\includegraphics[width=0.038\textwidth]{Legend_2}
		\caption{Two-dimentional oscillation patterns of $\sqrt[6]{|\cal S|}$ distributions at fixed total energy ($E_q=4\,{\rm MeV}$) for Laguerre-Gaussian vortex states collision. The distribution has been normalized so that its biggest value is $1$. $(\ell_1,\,\ell_2)$ is shown in every sub-picture.  Other parameters of initial states: $\bar k_{z1}=\bar k_{z2}=2 \,{\rm MeV},\,\sigma_{z1}=\sigma_{z2}=\sigma_{\perp 1}=\sigma_{\perp 2}=0.1\,{\rm MeV},\,n_1=n_2=0$.}\label{LG oscillation}
	\end{figure}\FloatBarrier

\end{document}